\documentclass[10pt,twocolumn]{article}
\usepackage{times}

\usepackage{amssymb}
\usepackage{amsmath}
\usepackage{amsfonts}
\usepackage{url, xspace} 
\usepackage[small,bf]{caption}
\usepackage{algorithm}
\usepackage{graphicx}
\usepackage{subfig}
\usepackage{bm}
\usepackage{soul, color}
\soulregister\cite7
\soulregister\ref7
\usepackage[noend]{algpseudocode}
\usepackage{enumitem}

\usepackage{hyperref}
\hypersetup{
  colorlinks=true,      
  linkcolor=blue,       
  citecolor=magenta,    
  filecolor=cyan,       
  urlcolor=red          
}

\usepackage{multirow}
\usepackage{hhline}

\usepackage{fnpos}

\newcommand{\reference}[2]{
	\ifthenelse{\boolean{isTechReport}}
	    {\ref{#1}} 
	    {#2}}

\newcommand{\myvec}[1]{\protect\overrightarrow{#1}}

\newcommand{\desc}[1]{}

\newcommand{\delete}[1]{}
\newcommand{\new}[1]{#1}

\usepackage{titlesec}
\titlespacing*{\section}{0pt}{0pt}{1.5pt}
\titlespacing*{\subsection}{0pt}{0pt}{1pt}
\titlespacing*{\subsubsection}{0pt}{0pt}{0.5pt}

\usepackage{etoolbox}
\makeatletter
\patchcmd{\ttlh@hang}{\parindent\z@}{\parindent\z@\leavevmode}{}{}
\patchcmd{\ttlh@hang}{\noindent}{}{}{}
\makeatother
\newcommand{\todo}[1]{}

\newcommand{\ignore}[1]{}

\def\name{{BoPF~}}

\def\ie{{i.e.}}
\def\eg{{e.g.}}

\makeatletter
\def\BState{\State\hskip-\ALG@thistlm}
\makeatother

\def\batchq{{TQ}}
\def\burstq{{LQ}}

\begin{document}

\title{\name: Mitigating the Burstiness-Fairness Tradeoff in Multi-Resource Clusters}

\author{Paper \# , \pageref{EndOfPaper}   Pages}

\author{Tan N. Le$^{1,2}$, Xiao Sun$^1$, Mosharaf Chowdhury$^3$, Zhenhua Liu$^1$ \\
\small {\em  $^1$ Stony Brook University \quad
	       $^2$ SUNY Korea \quad 
          $^3$ University of Michigan} \\ [2mm]
}
\date{}
\maketitle

\begin{abstract}
Simultaneously supporting latency- and throughout-sensitive workloads in a shared environment is an increasingly more common challenge in big data clusters. 
Despite many advances, existing cluster schedulers force the same performance goal -- fairness in most cases -- on all jobs. 
Latency-sensitive jobs suffer, while throughput-sensitive ones thrive. 
Using prioritization does the opposite: it opens up a path for latency-sensitive jobs to dominate. 
In this paper, we tackle the challenges in supporting both short-term performance and long-term fairness simultaneously with high resource utilization by proposing Bounded Priority Fairness (\name). 
\name provides short-term resource guarantees to latency-sensitive jobs and maintains long-term fairness for throughput-sensitive jobs. 
\name is the first scheduler that can provide long-term fairness, burst guarantee, and Pareto efficiency in a strategyproof manner for multi-resource scheduling. 
Deployments and large-scale simulations show that \name closely approximates the performance of Strict Priority as well as the fairness characteristics of DRF. 
In deployments, \name speeds up latency-sensitive jobs by $5.38\times$ compared to DRF, while still maintaining long-term fairness. 
In the meantime, \name improves the average completion times of throughput-sensitive jobs by up to $3.05\times$ compared to Strict Priority.

\end{abstract}

\section{Introduction}

\begin{figure}[!t]
  \centering
  \includegraphics[width=0.9\columnwidth]{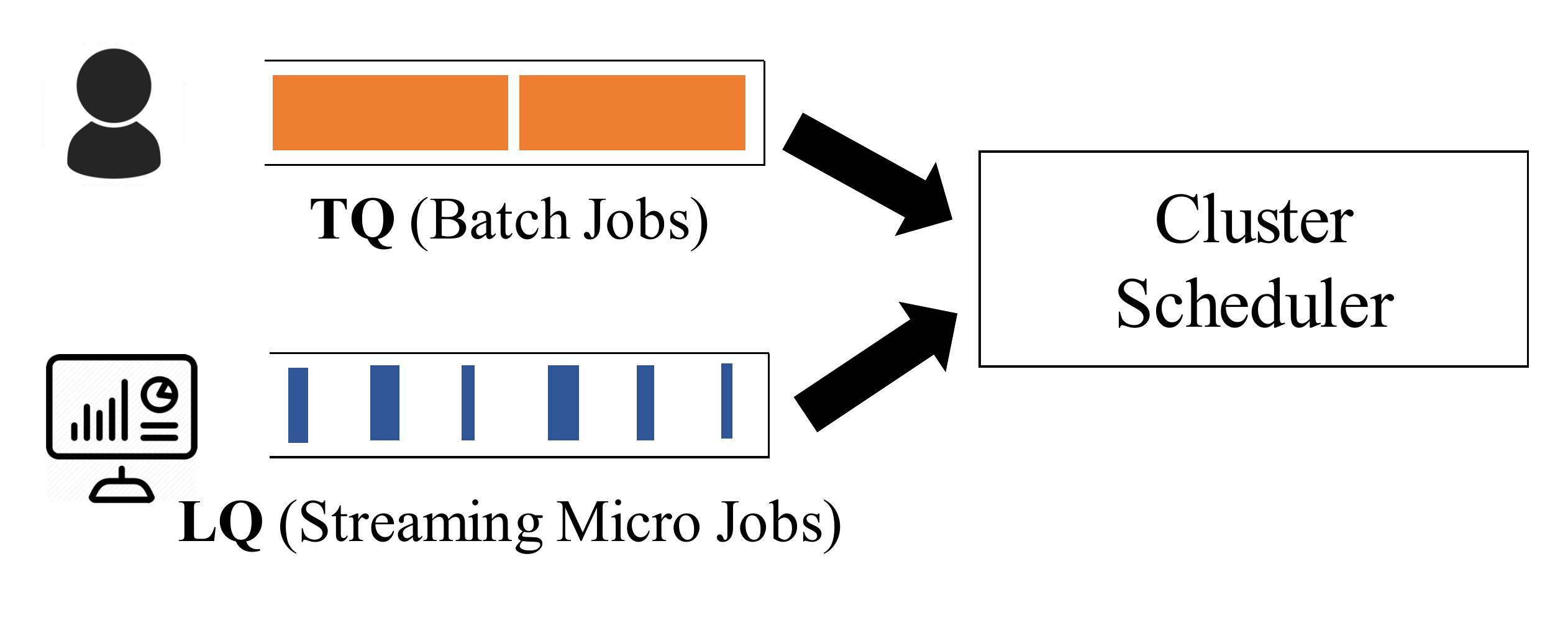}%
  \caption{Users and automated processes submit throughput-sensitive (TQ) and latency-sensitive (LQ) to the same cluster.}%
  \label{fig:queues}
\end{figure}

Cloud computing infrastructures are increasingly being shared between diverse workloads with heterogeneous resource requirements. 
In particular, throughput-sensitive batch processing systems \cite{mapreduce, dryad, graphlab, tez} are often complemented by latency-sensitive interactive analytics \cite{spark, blinkdb, presto} and online stream processing systems \cite{spark-streaming, trident, millwheel, naiad, flink}.
Simultaneously supporting these workloads is a balancing act between distinct performance metrics. For example, Figure~\ref{fig:queues} shows a cluster scheduling a mix of jobs from a throughput-sensitive queue (TQ) and a latency-sensitive queue (LQ).
Batch processing workloads such as indexing \cite{mapreduce} and log processing \cite{hadoop, spark} may submit hours-long large jobs via TQ. 
The \emph{average} amount of resources received over a certain period of time is critical for these jobs.
In contrast, interactive \cite{spark, presto} and online streaming \cite{spark-streaming, trident} workloads respectively submit on-demand and periodic smaller jobs via LQ. 
Therefore, receiving enough resources immediately for an LQ job is more important than the average resources received over longer time intervals.

To address the diverse goals, today's schedulers are becoming more and more complex.
They are multi-resource \cite{drf, tetris, multiresource-mungchiang, orchestra, pacman}, DAG-aware \cite{aalo, tetris, spark}, and allow a variety of constraints \cite{late, quincy, mantri, choosy, delay-scheduling}. 
Given all these inputs, they optimize for objectives such as fairness \cite{drf, jaffe-maxmin, drfq, hdrf}, performance \cite{sjf}, efficiency \cite{tetris}, or different combinations of the three \cite{carbyne, graphene}.
However, \new{most} existing schedulers have one shortcoming in common: \emph{they force the same performance goal on all jobs while jobs may have distinct goals}; therefore, they fail to provide performance guarantee in the presence of multiple types of workloads with different performance metrics. 
In fact, the performance of existing schedulers can be \emph{arbitrarily bad} for some workloads.

\begin{figure}[!t]
	\centering
	\includegraphics[width=0.2\linewidth]{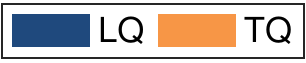} 
	\\
	\subfloat[DRF ensures instantaneous fairness, but increases the completion times of latency-sensitive jobs.]
  {\includegraphics[width=0.8\linewidth]{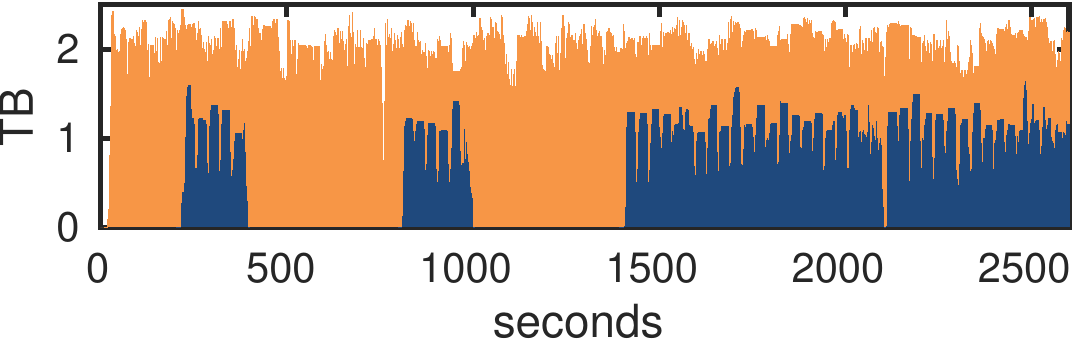}\label{fig:motiv-DRF}}
	\\ 
	\subfloat[SP decreases completion times of latency-sensitive jobs, but batch jobs do not receive their fair shares.]
  {\includegraphics[width=0.8\linewidth]{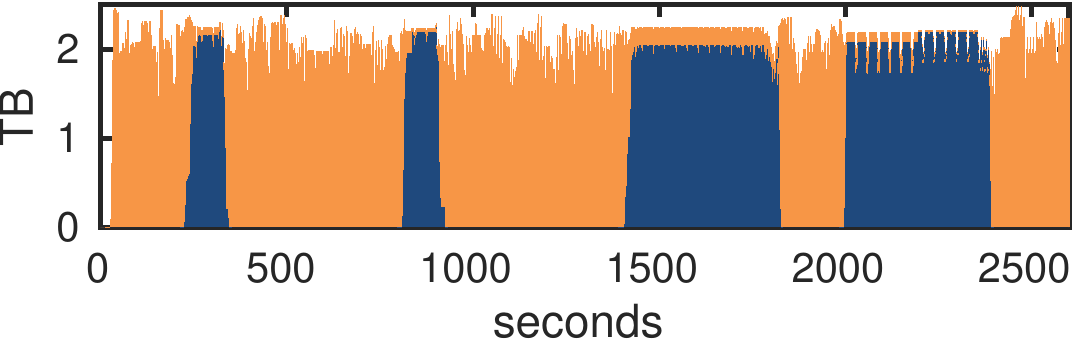}\label{fig:motiv-Strict}}
	\\ 
	\subfloat[The ideal solution allows first two latency-sensitive jobs to finish as quickly as possible, but protects batch jobs from latter \burstq jobs by ensuring long-term fairness.]
  {\includegraphics[width=0.8\linewidth]{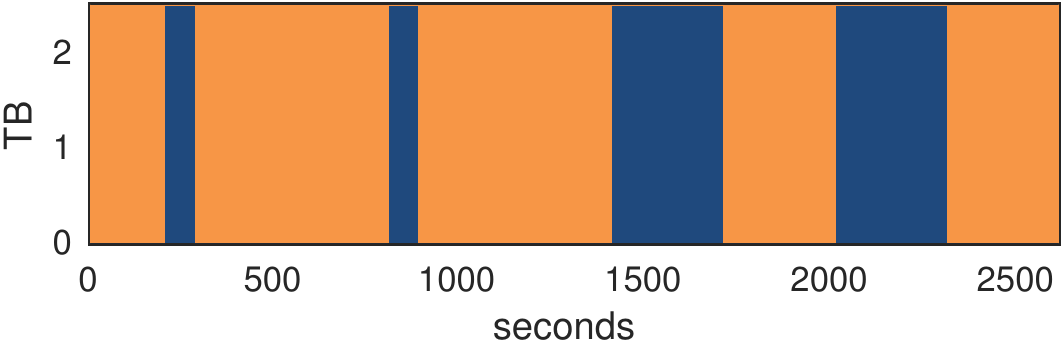}\label{fig:motiv-optimal}}	
  
	\caption{Need for bounded priority and long-term fairness in a shared multi-resource cluster with latency-sensitive (\burstq: blue/dark) and throughput-sensitive (\batchq: orange/light) jobs. The blank part on the top is due to resource fragmentation and overheads in Apache YARN.
    Although we focus only on memory allocations here, similar observations hold in multi-resource scenarios.}
	\label{fig:motiv_ex}
		 
\end{figure}

Consider the simple example in Figure~\ref{fig:motiv_ex} that illustrates the inefficiencies of existing resource allocation policies, in particular, DRF~\cite{drf} and Strict Priority (SP) \cite{strict_priority} for coexisting workloads with different requirements. 
In this example, we run memory-bound jobs in 
a 40-node cluster, where each node has 32 CPU cores and 64 GB RAM.
There are two queues, where each queue contains a number of jobs with the same performance goals.  
Apache Hadoop YARN \cite{hadoop-fair-scheduler} is set up to manage the resource allocation among the queues.
The first queue is for Spark streaming, which submits a simple MapReduce-like job every 10 minutes. 
We call it latency-sensitive queue (\burstq) because it aims to finish each of the jobs as quickly as possible.
The second queue is a throughout-sensitive batch-job queue (\batchq) formed by jobs generated from the BigBench workload~\cite{bbench} and queued up at the beginning. \batchq cares more about its long-term averaged resources received, e.g., every 10 minutes.
For simplicity of exposition, all jobs are memory-bound. 
%
We consider two common classes of policies -- priority-based and fairness-based allocation -- in this example, where the former is optimized for latency and the latter for fairness.
The memory resource consumption
under these two policies is depicted in Figures~\ref{fig:motiv-DRF} and~\ref{fig:motiv-Strict}, respectively. 
We defer the discussion of other policies to Section~\ref{sec:property-analysis}. 

SP gives {\burstq} the whole cluster's resources (high priority) whenever it has jobs to run; hence, it provides the lowest possible response time. 
For the first two arrivals, the average response time is 130 seconds. 
A detrimental side effect of SP, however, is that there is no resource isolation -- {\batchq} jobs may not receive any resources at all! 
In particular, {\burstq} is incentivized to increase its arrival rate -- e.g., for more accurate sampling and more iterations in training neural networks -- without any punishment. 
As it does so from the third job arrival, {\batchq} no longer receives its fair share. 
In the worst case, {\burstq} can take all the system resources and \emph{starve} {\batchq}. 
In summary, SP provides the best response time for {\burstq} , but no performance isolation for {\batchq} at all.
In addition, SP is incapable of handling multiple {\burstq}s. 

In contrast, DRF enforces \emph{instantaneous} fair allocation of resources at all times. 
During the burst of {\burstq}, {\burstq}  and {\batchq} share the bottleneck resource (memory) evenly until the jobs from {\burstq}  complete; then {\batchq} gets all resources before the next burst of {\burstq}. 
Clearly, {\batchq} is happy at the cost of longer completion times of \burstq's jobs, whose response time increases by 1.6 times. 
In short, DRF provides the best performance isolation for {\batchq}, but no performance consideration for {\burstq}. 
When there are many {\batchq}s, the response time of {\burstq} can be very large. 

Clearly, it is impossible to achieve the best response time under \emph{instantaneous} fairness. 
In other words, there is a hard tradeoff between providing instantaneous fairness for {\batchq}s and minimizing the response time of {\burstq}s.
Consequently, we aim to answer the following fundamental question in this paper: \emph{how well can we simultaneously accommodate multiple classes of workloads with performance guarantees, in particular, performance isolation for {\batchq}s and low response times for {\burstq}s}? 

\begin{figure}[!t]
  \centering
  \includegraphics[width=1.0\linewidth]{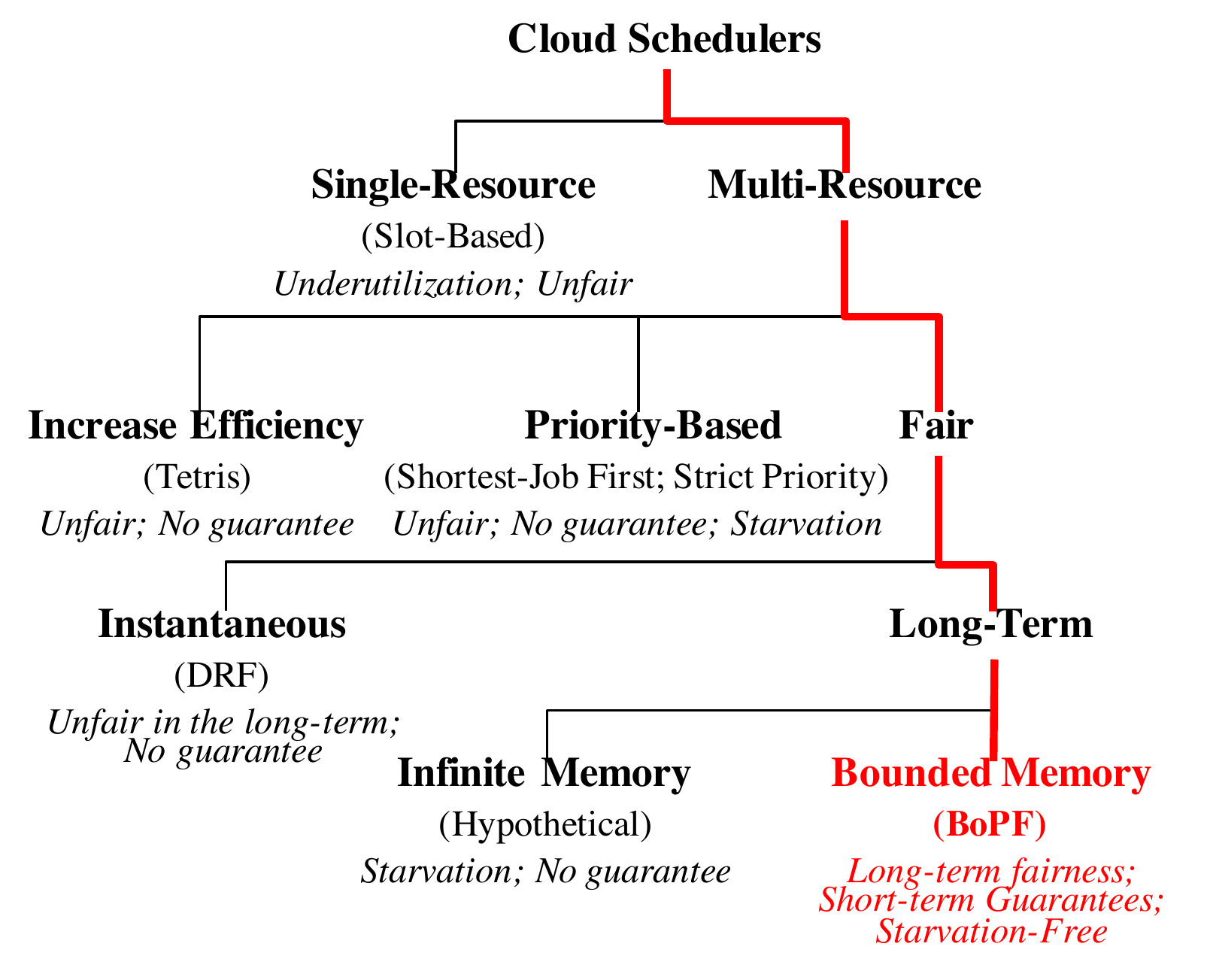}%
  \caption{{\name} in the cluster scheduling design space.}%
  \label{fig:design-space}
\end{figure}

We answer this question by designing \name: the first multi-resource scheduler that achieves both performance isolation for {\batchq}s in terms of \emph{long-term} fairness and response time guarantees for {\burstq}s, and is strategyproof. 
It is simple to implement and provides significant performance improvements even in the presence of uncertainties. 
The key idea is ``bounded'' priority for {\burstq}s: as long as the burst is not too large to hurt the long-term fair share of {\batchq}s, they are given higher priority so jobs can be completed as quickly as possible. 
Figure~\ref{fig:design-space} shows \name in the context of cluster scheduling landscape. 

We make the following contributions in this paper.

\textbf{Algorithm design.} We develop \name with the rigorously proven properties of strategyproofness, short-term bursts, long-term fairness, and high system utilization (\S\ref{sec:approach}). When {\burstq}s have different demands for each arrival, we further design mechanism to handle the uncertainties.


\textbf{Design and implementation.} 
We have implemented \name on Apache YARN \cite{yarn} (\S\ref{sec:impl}).
Any framework that runs on YARN can take advantage of \name.
The \name scheduler is implemented as a new scheduler in Resource Manager that runs on the master node.
The scheduling overheads for admitting queues or allocating resources are negligibly less than 1 ms for 20,000 queues.

%

\textbf{Evaluation on both testbed experiments and large-scale simulations.}
In deployments, \name provides up to $5.38\times$ lower completion times for {\burstq} jobs than DRF, while maintaining the same long-term fairness (\S\ref{sec:testbed}). 
At the same time, \name provides up to $3.05\times$ more fair allocation to {\batchq} jobs compared to SP.

\section{Motivation}
\label{sec:motivation}
%



\subsection{Benefits of Temporal Co-Scheduling}
\label{sec:ex}

Consider the example in Figure~\ref{fig:motiv_ex} again. Recall that SP and DRF are two extreme cases in trading off performance and fairness: SP provides the best performance (for {\burstq}s) with no fairness consideration (for {\batchq}s); DRF ensures the best isolation (for {\batchq}s) with poor performance (of {\burstq}s). 
However, it is still possible for {\burstq}s and {\batchq}s to share the cluster by thoughtful co-scheduling over time.

The ideal allocation is depicted in Figure~\ref{fig:motiv-optimal}. 
The key idea is ``bounded'' priority for {\burstq}s as we discussed in the previous section. 
In particular, before 1,400 seconds, \burstq's bursts are small, so it gets higher priority, which is similar to SP. 
After {\burstq} increases its demand, only a fraction of its demand can be satisfied with the entire system's resources. Then it has to give resources back to {\batchq} to ensure long-term fairness.

\subsection{Desired Properties}
\label{sec:desire}


We restrict our attention in this paper to the following, important properties: burst guarantee for {\burstq}s, long-term fairness for {\batchq}s, strategyproofness, and Pareto efficiency to improve cluster utilization.

\textbf{Burst guarantee (BG)} provides performance guarantee for {\burstq}s by allocating guaranteed amount of resources during their bursts. 
In particular, an {\burstq} requests its minimum required resources for its bursts to satisfy its service level agreements, e.g., percentiles of response time. 


\textbf{Long-term fairness (LF)} provides every queue in the system the same amount of resources over a (long) period, e.g., 10 minutes.
Overall, it ensures that {\batchq}s progress no slower than any {\burstq} in the long run. LF implies sharing incentive, which requires that each queue should be better off sharing the cluster, than exclusively using its own static share of the cluster. 
If there are $n$ queues, each queue cannot exceed $\frac{1}{n}$ of all resources under static sharing.\footnote{For simplicity of presentation, we consider queues with the same weights, which can be easily extended to queues with different weights.}

\textbf{Strategyproofness (SPF)} ensures that queues cannot benefit by lying about their resource demands. 
This provides incentive compatibility, as a queue cannot improve its allocation by lying. 

\textbf{Pareto efficiency (PE)} is about the optimal utilization of the system. 
A resource allocation is Pareto efficient if it is impossible to increase the allocation/utility of a queue without hurting at least another queue. 

\subsection{Analysis of Existing Policies}
\label{sec:property-analysis}

\begin{table}[!t]
	\small
	\centering
	
	\begin{tabular}{ |c||c|c|c||c| } 
		\hline
		Property  & SP  & DRF  & M-BVT  & \name \\   [0.5ex] 
		\hline\hline
		Burst Guarantee (BG)    &  \checkmark* &$\times$ & \checkmark*  &\checkmark \\ 
		Long-Term Fairness (LF) & $\times$ & \checkmark &\checkmark &  \checkmark \\
		Strategyproofness (SPF)  & $\times$ & \checkmark& $\times$ & \checkmark\\
		Pareto Efficiency (PE)  & \checkmark & \checkmark &\checkmark & \checkmark \\
		\hline \hline 
		Single Resource Fairness & $\times$  & \checkmark & \checkmark & \checkmark \\
		Bottleneck Fairness & $\times$  & \checkmark & \checkmark & \checkmark \\
		Population Monotonicity & \checkmark  & \checkmark & \checkmark & \checkmark \\
		\hline
	\end{tabular}
	\caption{Properties of existing policies and {\name}. $\checkmark^*$ means that the property holds when there is only one \burstq.}
	\label{tab-properties}
	
\end{table}


\textbf{Strict Priority (SP): }
SP is employed to provide performance guarantee for {\burstq}s. As the name suggests, an SP scheduler always prioritize {\burstq}s. Therefore, when there is only one \burstq, SP provides the best possible performance guarantee. 
However, when there are more than one {\burstq}s, it is impossible to give all of them the highest priority. Meanwhile, {\batchq}s may not receive enough resources, which violates long-term fairness. As the {\burstq}s may request more resources than what they actually need, strategyproofness is not enforced, and therefore the system may waste some resources -- i.e., it is not Pareto efficient. 

%

\textbf{DRF:}  
DRF is an extension of max-min fairness to the multi-resource environment, where the dominant share is used to map the resource allocation (as a vector) to a scalar value. 
It provides instantaneous fairness, strategyproofness, and Pareto efficiency. 
However, because DRF is an instantaneous allocation policy without any memory, it cannot prioritize jobs with more urgent deadlines. In particular, no burst guarantee is provided. 
Even assigning queues different weights in DRF is homogeneous over time and cannot provide the burst guarantee needed. In addition, there is no admission control. Therefore, as the number of queues increases, no queue's performance can be guaranteed.


\textbf{M-BVT: }
BVT~~\cite{bvt} \delete{was designed} \new{is a thread-based CPU scheduler} for a mix of real-time and best-effort tasks. The idea is that for real-time tasks, BVT allows them to borrow some virtual time (and therefore resources) from the future and be prioritized for a period without increasing their long-term shares. 

\delete{To make it comparable, we extend the idea of it to M-BVT for a multi-resource environment} 
\new{Since BVT was designed for a single-resource environment, we extend the idea of BVT to M-BVT for multiple resources.}
Under the M-BVT policy, \burstq-$i$ is assigned a virtual time warp parameter $W_i$, which represents the urgency of the queue. Upon an arrival of its burst at $A_i$, an effective virtual time $E_i = A_i-W_i$ is calculated. This is used as the priority (smaller $E_i$ means higher priority) for scheduling. When \burstq-$i$ has the only smallest $E_i$, it may use the whole system's resources and its $E_i$ increases at the rate of its progress calculated by DRF. Eventually, its $E_i$ is no longer the only smallest. Then resources are shared in a DRF-fashion among queues with the smallest virtual times.


M-BVT has some good properties. For instance, the DRF component ensures long-term fairness, and the BVT component strives for performance. Pareto efficiency follows from the work conservation of the policy. 

However, it does not provide general burst guarantees as any new arriving queue (with larger virtual time warp parameter) may occupy the resources of existing {\burstq}s or share resources with them, thus hurting their completion time. In addition, it is not strategyproof because queues can lie about their needs in order to get a larger virtual time warp.



\textbf{Other policies} like the CEEI~\cite{moulin2014cooperative} provide fewer desired properties. 

\subsection{Summary of the Tradeoffs}

As listed in Table~\ref{tab-properties}, no prior policy can simultaneously provide all the desired properties of fairness/isolation for {\batchq}s while providing burst guarantees for all the {\burstq}s with strategyproofness. In particular, if strict priority is provided to an \burstq without any restriction for its best performance (e.g., strategyproofness), there is no isolation protection for {\batchq}s' performance. On the other hand, if the strictly instantaneous fairness is enforced (e.g., DRF), there is no room to prioritize short-term bursts. While the idea in M-BVT is reasonable, it is not strategyproof and cannot provide burst guarantee.


The key question of the paper is, therefore, how to allocate system resources in a near-optimal way; meaning, satisfying all the critical properties in Table~\ref{tab-properties}.

\section{\name: A Scheduler With Memory}
\label{sec:approach}

In this section, we first present the problem setting (\S\ref{sec:settings}) and then formally model the problem in Section~\ref{sec:properties}. \name achieves the desired properties by admission control, guaranteed resource provision, and spare resource allocation (\S\ref{sec:solution_approach}). 
Finally, we prove that BPF satisfies all the properties in Table~\ref{tab-properties} (\S\ref{sec:properties-theorem}).

\subsection{Problem Settings}\label{sec:settings}

We consider a system with $K$ resources. The capacity of resource $k$ is denoted by $C^k$. The system resource capacity is therefore a vector $\myvec{C} = \langle C^1, C^2, ..., C^K \rangle.$
For ease of exposition, we assume $\myvec{C}$ is a constant over time, but our methodology applies directly to the cases with time-varying $\myvec{C}(t)$, e.g., with estimations of $\myvec{C}(t)$ at the beginning and leveraging stochastic optimization \cite{schneider2007stochastic} and online algorithm design \cite{jaillet2012online}.

We restrict our attention to {\burstq}s for interactive sessions and streaming applications, and {{\batchq}}s for batch jobs. 

{\burstq}-$i$'s demand comes from a series $N_i$ of bursts, each consisting of a number of jobs. We denote by $T_i(n)$ the arrival time of the $n$-th burst, which must be finished within $t_i(n)$. Therefore, its $n$-th burst needs to be completed by $T_i(n)+t_i(n)$ (i.e., deadline).
Denote the demand of its $n$-th arrival by a vector $\myvec{d_i}(n)=\langle d^1_i(n), d^2_i(n), ..., d^K_i(n)\rangle$, where $d^k_i(n)$ is the demand on resource-$k$.

In practice, inter-arrival time between consecutive bursts $T_i(n+1)-T_i(n)$ can be fixed for some applications such as Spark Streaming \cite{spark-streaming}, or it may vary for interactive user sessions. In general, the duration is quite short, e.g., several minutes. 
Similarly, the demand vector $\myvec{d_i}(n)$ may contain some uncertainties, and we assume that queues have their own estimations. Therefore, our approach has to be strategyproof so that queues report their estimated demand, 
as well as their true deadlines. 


To enforce the long-term fairness, the total demand of {\burstq}-$i$'s $n$-th arrival $\myvec{d_i}(n)$ should not exceed its fair share, which can be calculated by a simple fair scheduler -- i.e., $\frac{\myvec{C}(T_i(n+1)-T_i(n))}{N}$, when there are $N$ queues admitted by {\name} -- or a more complicated one such as DRF. We adopt the former in analysis because it provides a more conservative evaluation of the improvements brought by {\name}. 


In contrast, {{\batchq}}'s jobs are queued at the beginning with much larger demand than each burst of {\burstq}s. 


\begin{table}[!t]
\small
\centering

\begin{tabular}{|c|l|} \hline
Notation & Description \\ \hline \hline
$\mathbb{H}$ & Admitted {\burstq}s with hard guarantee \\ \hline
$\mathbb{S}$ & Admitted {\burstq}s with soft guarantee \\ \hline
$\mathbb{E}$ & Admitted {\batchq}s and {\burstq}s with fair share only \\ \hline
\hline\end{tabular}

\caption{Important notations} 
 
\label{tbl:notations}
\end{table}


\subsection{Modeling the Problem}
\label{sec:properties}


\textbf{Completion time:}
Let us denote by $R_i(n)$ the (last) completion time of jobs during {\burstq}-$i$'s $n$-th arrival. If {\burstq}-$i$ is admitted with \emph{hard} guarantee, we ensure that a large fraction $\alpha_i$  of arrivals are completed before deadlines\footnote{$\alpha_i$ can be 95\% or 99\% depending on the SLA.}; i.e., $\sum_{n\in N_i}\mathbf{1}_{\left\{R_i(n)\leq T_i(n)+t_i(n)\right\}} \geq \alpha_i|N_i|$, where $\mathbf{1}_{\left\{\cdot\right\}}$ is the indicator function which equals to 1 if the condition is satisfied and 0 otherwise, $|N_i|$ is the number of arrivals of {\burstq}-$i$. 
\new{A more general function is considered as the future work.}
If {\burstq}-$i$ is admitted with only \emph{soft/best-effort} guarantee, we maximize the fraction of arrivals completed on time.

\textbf{Long-term fairness:}
Denote by $\myvec{a_i}(t)$ and $\myvec{e_j}(t)$ the resources allocated for {\burstq}-$i$ and {\batchq}-$j$ at time $t$, respectively.
For a possibly long evaluation interval $[t, t+T]$ during which there is no new admission or exit, the average resource guarantees received are calculated as $\frac{1}{T}\int_t^{t+T} \myvec{a_i}(\tau)\text{d}\tau$ and $\frac{1}{T}\int_t^{t+T} \myvec{e_j}(\tau)\text{d}\tau$.
We require the allocated dominant resource, i.e., the largest amount of resource allocated across all resource types, received by any {\batchq} queue is no smaller than that received by an {\burstq}. Formally, $\forall i\in\mathbb{A}, \forall j\in\mathbb{B}$, where $\mathbb{A}$ and $\mathbb{B}$ is the set of admitted {\burstq}s and {\batchq}s, respectively, 
\begin{small}$\max_{k}\left\{\frac{1}{T}\int_t^{t+T} a_i^k(\tau)\text{d}\tau\right\}\leq\max_{k}\left\{\frac{1}{T}\int_t^{t+T} e_j^k(\tau)\text{d}\tau\right\}$\end{small}, 
where $a_i^k(\tau)$ and $e_j^k(\tau)$ are allocated type-$k$ resources for {\burstq}-$i$ and {\batchq}-$j$ at time $\tau$, respectively.
This condition provides long-term protections for admitted {\batchq}s.

%

\textbf{The optimization problem:}
We would like to maximize the arrivals completed before the deadlines for admitted {\burstq}s with soft guarantee while meeting the specified fraction of deadlines of admitted {\burstq}s with hard guarantees and keeping the long-term fairness.

The decisions to be made are (i) admission control, which decides the set of admitted {\burstq}s ($\mathbb{H}$, $\mathbb{S}$) and the set of admitted {\batchq}s ($\mathbb{E}$); and (ii) resources allocated to admitted queues {\burstq}-$i$ and {\batchq}-$j$ ($\myvec{a_i}(t)$ and $\myvec{e_j}(t)$, respectively) over time. 
If there are some unused/unallocated resources, queues with unsatisfied demand can share them.


\subsection{Solution Approach}
\label{sec:solution_approach}

Our solution \name consists of three major components: admission control procedure to decide $\mathbb{H}$, $\mathbb{S}$ and $\mathbb{E}$, guaranteed resource provisioning procedure for $\myvec{a_i}(t)$, and a spare resource allocation procedure.

\textbf{Admission control procedure:} 
\name admits queues into the following three classes: 
\begin{itemize}[noitemsep,nolistsep]
 \item $\mathbb{H}$: {\burstq}s admitted with hard resource guarantee.
 \item $\mathbb{S}$: {\burstq}s admitted with soft resource guarantee. Similar to hard guarantee, but need to wait when some {\burstq}s with hard guarantee are occupying system resources. 
 \item $\mathbb{E}$: Elastic queues that can be either {\burstq}s or {\batchq}s. There is no burst guarantee, but long-term fair share is provided.
\end{itemize}

\begin{algorithm} [!ht]
	\small
\caption{\name Scheduler}
\label{algorithm1}
\begin{algorithmic}[1]
\Procedure{periodicSchedule()}{}
\If{there are new {\burstq}s $\mathbb{Q}$}
	\State $\{\mathbb{H},\mathbb{S},\mathbb{E}\}$ =\textsc{LQAdmit}($\mathbb{Q}$)\EndIf
\If{there are new {\batchq}s $\mathbb{Q}$}
	\State $\{\mathbb{E}\}$ =\textsc{TQAdmit}($\mathbb{Q}$)\EndIf
\State \textsc{allocate}($\mathbb{H},\mathbb{S}, \mathbb{E}$)
\EndProcedure
\\
\Function{LQAdmit({\burstq}s $\mathbb{Q}$)}{}
\ForAll{\burstq  $Q \in \mathbb{Q}$}	
\If{safety condition \eqref{eqn:ad-safety} satisfied}
	\If{fairness condition \eqref{eqn:ad-fair} satisfied}
		\If{resource condition \eqref{eqn:ad-enough} satisfied}
			\State Admit $Q$ to hard guarantee $\mathbb{H}$
		\Else
			\State Admit $Q$ to soft guarantee $\mathbb{S}$
		\EndIf
	\Else
		\State Admit $Q$ to elastic $\mathbb{E}$ with long-term fair share

	\EndIf
\Else 
	\State Reject $Q$
\EndIf
\EndFor
\State \textbf{return} $\{\mathbb{H},\mathbb{S},\mathbb{E}\}$	
\EndFunction
\\
\Function{TQAdmit(queue $\mathbb{Q}$)}{}
\ForAll{\batchq  $Q \in \mathbb{Q}$}	
	\If {safety condition \eqref{eqn:ad-safety} satisfied}
		\State Admit $Q$ to elastic $\mathbb{E}$ with long-term fair share
	\Else 
		\State Reject $Q$
	\EndIf
\EndFor
\State \textbf{return} $\{\mathbb{E}\}$
\EndFunction
\\
\Function{allocate($\mathbb{H}$, $\mathbb{S}$, $\mathbb{E}$)}{}
\ForAll{\burstq $Q \in \mathbb{H}$}
\State $\myvec{a_i}(t)=\frac{\myvec{d_i}(n)}{t_i(n)}$ for $t\in[T_i(n),T_i(n+1)]$ 
\EndFor
\ForAll{\burstq $Q \in \mathbb{S}$}
\State allocate $\myvec{C}-\sum_{j\in \mathbb{H}}\myvec{a_j}(t)$ based on SRPT until each {\burstq}-$i$'s allocation reaches $\myvec{d_i}(n)$ or the deadline arrives.
\EndFor
\State Obtain the remaining resources $\myvec{L}$
\State DRF($\mathbb{E}$, $\myvec{L}$)
\EndFunction

\end{algorithmic}
\end{algorithm}


\new{The system expects to admit at least $N_{min}$ queues.}
Before admitting \burstq-$i$, \name checks if admitting it invalidates any resource guarantees committed for {\burstq}s in $\mathbb{H}\cup\mathbb{S}$, i.e., the following \emph{safety condition} needs to be satisfied: 
\begin{align}
    \small
	\label{eqn:ad-safety}
    \begin{aligned}
	& \myvec{d_j}(n) \leq \frac{\myvec{C}\left(T_j(n+1)-T_j(n)\right)}{\max(|\mathbb{H}|+|\mathbb{S}|+|\mathbb{E}|+1, N_{min})}, \forall n, \forall j \in \mathbb{H}\cup\mathbb{S},
    \end{aligned}  
\end{align}
where $|\mathbb{H}|+|\mathbb{S}|+|\mathbb{E}|$ is the number of already admitted queues. 
If (\ref{eqn:ad-safety}) is not satisfied, \burstq-$i$ is rejected. Otherwise, it is safe to admit \burstq-$i$ and the next step is to decide which of the three classes it should be added to.

For \burstq-$i$ to have some resource guarantee, either hard or soft, its own total demand should not exceed its long-term fair share. Formally, the \emph{fairness condition} is 
\begin{align}
	\label{eqn:ad-fair}
    \begin{aligned}
	& \myvec{d_i}(n) \leq \frac{\myvec{C}\left(T_i(n+1)-T_i(n)\right)}{\max(|\mathbb{H}|+|\mathbb{S}|+|\mathbb{E}|+1, N_{min})}, \forall n, 
    \end{aligned}   
\end{align}

If only condition~(\ref{eqn:ad-safety}) is satisfied but (\ref{eqn:ad-fair}) is not, \burstq-$i$ is added to $\mathbb{E}$.
If both conditions~(\ref{eqn:ad-safety}) and (\ref{eqn:ad-fair}) are satisfied, it is safe to admit \burstq-$i$ to $\mathbb{H}$ or $\mathbb{S}$. If there are enough uncommitted resources (\emph{resource condition} (\ref{eqn:ad-enough})), \burstq-$i$ is admitted to $\mathbb{H}$. Otherwise it is added to $\mathbb{S}$.
\begin{align}
    \small 
	\label{eqn:ad-enough}
    \begin{aligned}
	& \frac{\myvec{d_i}(n)}{t_i(n)} \leq \myvec{C}-\sum_{j\in \mathbb{H}}\myvec{a_j}(t), \forall n, t\in[T_i(n), T_i(n)+t_i(n)]
    \end{aligned} 
\end{align}

%
%

For {\batchq}-$j$, \name simply checks the safety condition (\ref{eqn:ad-safety}). If it is satisfied, {\batchq}-j is added to $\mathbb{E}$. Otherwise {\batchq}-$j$ is rejected.

\textbf{Guaranteed resource provisioning procedure}
For each {\burstq}-$i$ in $\mathbb{H}$, during $[T_i(n),T_i(n)+t_i(n)]$, \name allocates constant resources to fulfill its demand $\myvec{a_i}(t)=\frac{\myvec{d_i}(n)}{t_i(n)}$. 
{\burstq}s in $\mathbb{S}$ shares the uncommitted resource $\myvec{C}-\sum_{j\in \mathbb{H}}\myvec{a_j}(t)$ based on SRPT (Shortest Remaining Processing Time)~\cite{bansal2001analysis} until each {\burstq}-$i$'s consumption reaches $\myvec{d_i}(n)$ or the deadline arrives.
\new{Meaning, BoPF prioritizes the LQs that are about to reach $\myvec{d_i}(n)$ before the deadline.}


After every {\burstq} in $\mathbb{H}$ and $\mathbb{S}$ is allocated, remaining resources are allocated to queues in $\mathbb{E}$ using DRF~\cite{drf}. 

%

%

\textbf{Spare resource allocation procedure}
If some allocated resources are not used, they are further shared by {\batchq}s and {\burstq}s with unsatisfied demand. This maximizes system utilization.

%
%
%
%

\subsection{Properties of \name}
\label{sec:properties-theorem}


First, we argue that \emph{\name ensures long-term fairness, burst guarantee, and Pareto efficiency}. 

The safety condition and fairness condition ensure the long-term fairness for all {\batchq}s. 

For {\burstq}s in $\mathbb{H}$, they have hard resource guarantee and therefore can meet their SLA. For {\burstq}s in $\mathbb{S}$, they have resource guarantee whenever possible, and only need to wait after {\burstq}s in $\mathbb{H}$ when there is a conflict. Therefore, their performance is much better than if they were under fair allocation policies. 


The addition of $\mathbb{S}$ allows more {\burstq}s to be admitted with resource guarantee and therefore increases the resources utilized by {\burstq}s. Finally, we fulfill spare resources with {\batchq}s, so system utilization is maximized, reaching Pareto efficiency.



\new{In addition, we prove that \emph{\name is weak-strategyproof}. Meaning, users have limited incentive to lie about their demand}.
The detail of proof is in Appendix \ref{proof:strategyproof}.

\subsection{Handling Uncertainties}
\label{sec:uncertainties}

\todo{very hard to understand - not happy}

In practice, arrivals of {\burstq}-$i$ may have different sizes, i.e., $\myvec{d_i}$ is not deterministic but instead has some probability distributions. Here we extend \name to handle this case.

We assume {\burstq}-$i$ knows its distributions, e.g., from historical data. In particular, it knows the cumulative probability distribution of each resource $k$, denoted by $F_{ik}$ if these distributions on multiple resources are independent. 
The requirement regarding $\alpha_i$ can be converted into $\prod_j F^{-1}_{ik}(d_{ik}) \geq \alpha_i$, where $d_{ik}$ is the request demand on resource $k$. This gives $F^{-1}_{ik}(d_{ik}) \geq \alpha_i^{1/K}$. Finally, the request on resource-$k$ $d_{ik}=F_{ik}(\alpha_i^{1/K})$. We call this $\alpha$-strategy.

When distributions of multiple resources are correlated, we only have the general form $F^{-1}_i (\myvec{d_i}) \geq \alpha_i$, where $F_{i}$ is the joint distribution on all resources. We have the following properties in this case.\footnote{The proof is omitted due to space limit.}
When the distributions are pairwise positively correlated, $\alpha$-strategy over-provisions resources. If they are pairwise negatively correlated, $\alpha$-strategy under-provisions resources. Numerical approaches can be applied to adjust the $\alpha$-strategy accordingly. 

Taking the correlations on multiple resources into consideration is important. In particular, when these distributions are perfectly correlated, $d_{ik}=F_{ik}(\alpha_i^{1/K})$ can be reduced to $d_{ik}=F_{ik}(\alpha_i)$.
\new{When the standard deviation is large (e.g., 40\% of the mean for Normal distribution), the demand can be reduced by 10\% with three resources.} This increases the chance of \burstq-$i$ being admitted.

\section{Design Details}
\label{sec:impl}

In this section, we describe how we have implemented \name on Apache YARN,
how we use standard techniques for demand estimation, and additional details related to our implementation.

\begin{figure}
\centering

\includegraphics[width=1.0\linewidth]{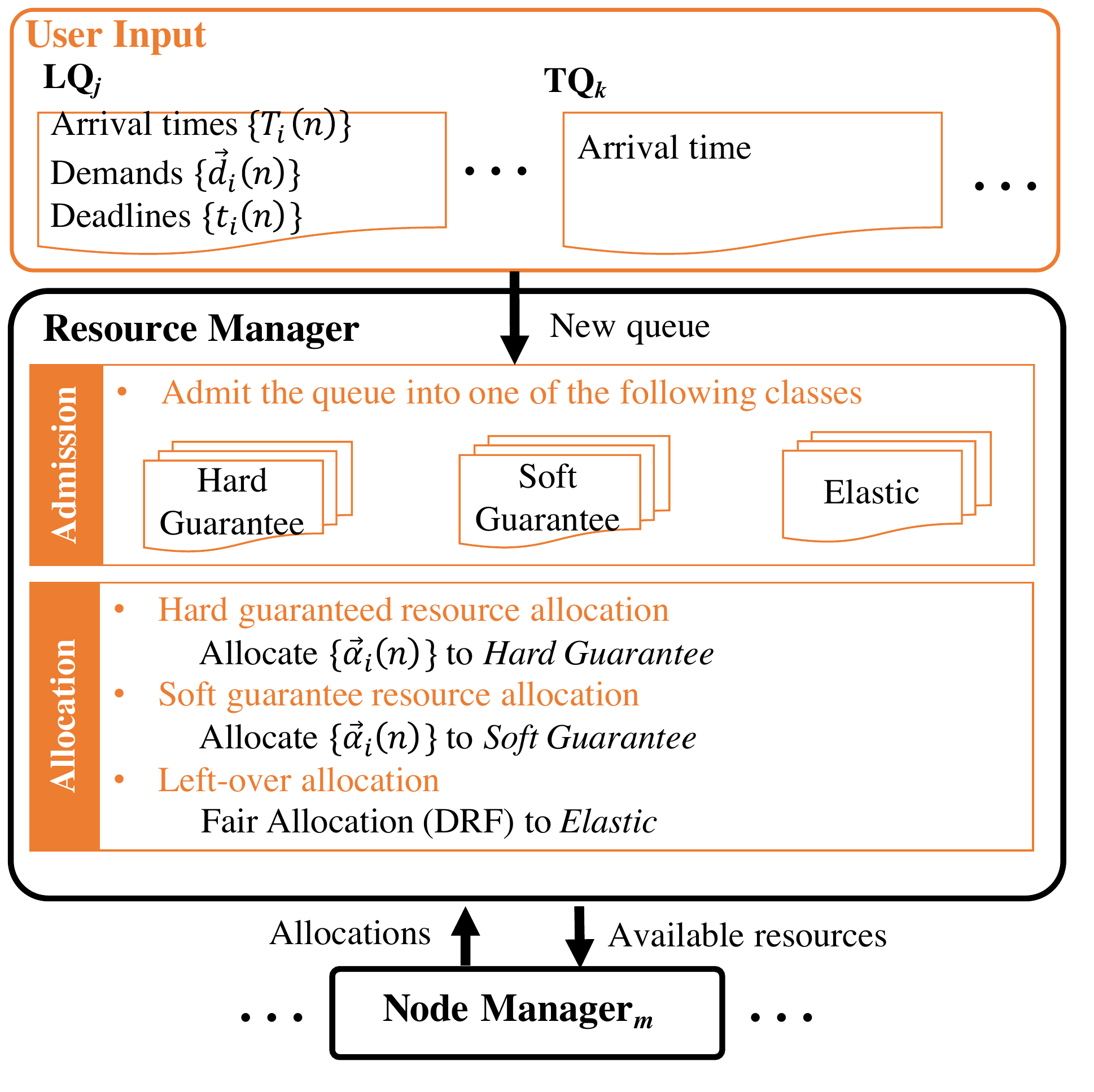}
 
\caption{Enabling bounded prioritization with long-term fairness in a  multi-resource cluster. 
\name-related changes are shown in orange.}
\label{fig:system_design}
 
\end{figure}

\subsection{Enabling \name in Cluster Managers}

Enabling bounded prioritization with long-term fairness requires implementing the \emph{\name scheduler} itself along with an additional \emph{admission control} module in cluster managers, and it takes \emph{additional information on demand characteristics} from the users. 
A key benefit of {\name} is its simplicity of implementation: we have implemented it in YARN.
In the following, we describe how and where we have made the necessary changes.

\textbf{Primer on Data-Parallel Cluster Scheduling}
Modern cluster managers typically includes three components: \emph{job manager} or application master (AM), \emph{node manager} (NM), and \emph{resource manager} (RM).

One NM runs on each server, which is responsible for managing resource containers on that server. 
A container is a unit of allocation and are used to run specific tasks. 

For each application, a job manager or AM interacts with the RM to request job demands and receive allocation and progress updates. 
It can run on any server in the cluster. 
AM manages and monitors job demands (memory and CPU) and job status (\texttt{PENDING}, \texttt{IN\_PROGRESS}, or \texttt{FINISHED}). 

The RM is the most important part in terms of scheduling. 
It receives requests from AMs and then schedules resources using an operator-selected scheduling policy. 
It asks NM to prepare resource containers for the various tasks of the submitted jobs.

\textbf{\name Implementation}
We made three changes for taking user input, performing admission control, and calculating resource shares -- all in the RM.
We do not modify NM and AM.
Our implementation also requires more input parameters from the users regarding the demand characteristics of their job queues. 
Figure~\ref{fig:system_design} depicts our design.


\textbf{User Input} Users submit their jobs to their queues. 
In our system, there are 2 queue types, i.e., {\burstq}s and {\batchq}s. 
We do not need additional parameters for {\batchq}s because they are the same as the conventional queues.
Hence, we assume that {\batchq}s are already available in the system. 
However, the \name scheduler needs additional parameters for LQs; namely, arrival times and demands.
\new{Since LQs prefer to have resource guarantee, it is necessary for them to report their own demands. The demand can be estimated by using an off-line estimator like Ernest \cite{ernest}. The estimation is not necessarily accurate. Nonetheless, we will show that our system is robust to large errors in Section~\ref{sec:estimation_error_sim}.}

A user submits requests containing their parameters of the new \burstq. 
After receiving the parameters, the RM sets up a new {\burstq} queue for the user.
Users can also ask the cluster administrator to set up the parameters.

\textbf{Admission Control} YARN does not support admission control. 
We implement an admission control module to classify {\burstq}s and {\batchq}s into Hard Guarantee, Soft Guarantee, and Elastic classes. 
A new queue is rejected if it cannot meet the safety condition \eqref{eqn:ad-safety}, which invalids the committed performance.
If it is a \batchq, it is added into the Elastic class.
If the new {\burstq} does not satisfy the fairness condition \eqref{eqn:ad-fair}, it is also admitted to the Elastic class.
If the new {\burstq} meets the fairness condition \eqref{eqn:ad-fair}, but fails at the resource condition \eqref{eqn:ad-enough}, it will be put in the Soft Guarantee class.
If the new {\burstq} meets all the three conditions, i.e., safety, fairness, and resource, it will be admitted to the Hard Guarantee class.

\textbf{\name Scheduler} We implement \name as a new scheduling policy to achieve our joint goals of bounded priority with long-term fairness. 
Upon registering the queues, users submit their jobs to their {\burstq}s or {\batchq}s. 
Thanks to admission control, {\burstq}s and {\batchq}s are classified into Hard Guarantee, Soft Guarantee, and Elastic classes. 
Note that resource sharing policies are implemented across queues in YARN, jobs in the same queue are scheduled in FIFO manner.
Hence, \name only sets the share at the individual queue level.

\name Scheduler periodically set the share levels to all {\burstq}s in Hard Guarantee and Soft Guarantee classes.
These share levels are upper-bounds on resource allocation that an {\burstq} can receive from the cluster. 
Based on the real demand of {\burstq}s, \name allocates resources until it meets the share levels\delete{ (whether it is in the ON period or OFF period)}. 

\name Scheduler allocates the resource to the three classes in the following priority order: (1) Hard Guarantee class, (2) Soft Guarantee class, and (3) Elastic class. 
The {\burstq}s in the Hard Guarantee class are allocated first.
Then, the \name continues allocates the resource to the {\burstq}s in Soft Guarantee class.
The queues in the Elastic class are allocated with leftover resources using DRF \cite{drf}.


\subsection{Demand Estimation}

\new{
\name requires accurate estimates of resource demands and their durations of {\burstq} jobs by users. 
These estimations can be done by using well-known techniques. For example, users can use history of prior runs \cite{rope, jockey, tetris} with the assumption that resource requirements for the tasks in the same stage are similar \cite{drf, pacman, paratimer}. The progress of pipelining jobs like SQL queries can be estimated by using the number of completed calls versus the total number of calls \cite{chaudhuri2004estimating,luo2004toward,konig2011statistical}.
For distributed jobs, their completion times given resource allocations can be estimated using machine learning techniques  \cite{paris, ernest, cherrypick}.
We do not make any new contributions on demand estimation in this paper. 
When LQs have bursty arrivals of different sizes, BPF with the $\alpha$-strategy ensures the performance with the average usage remains similar (\S\ref{sec:performance_large_scale}). 
We consider a more thorough study an important future work.}

\subsection{Operational Issues}

\textbf{Container Reuse}
Container reuse is a well-known technique that is used in some application frameworks, such as Apache Tez. 
The objective of container reuse is to reduce the overheads of allocating and releasing containers. 
The downside is that it causes resource waste if the container to be reused is larger than the real demand of the new task. 
Furthermore, container reuse is not possible if the new task requires more resource than existing containers. 
For our implementation and deployment, we do not enable container reuse because {\name} periodically prefers more free resources for {\burstq} jobs, causing its drawbacks to outweigh its benefits.

\textbf{Preemption}
Preemption is a recently introduced setting in the YARN Fair Scheduler \cite{hadoop-fair-scheduler}, and it is used to kill running containers of one job to create free containers for another. 
By default, preemption is not enabled in YARN. 
For {\name}, using preemption can help in providing guarantees for {\burstq}s. 
However, killing the tasks of running jobs often results in failures and significant delays. 
We do not use preemption in our system throughout this paper.

\section{Evaluation}

We evaluated \name using three widely used big data benchmarks -- BigBench (BB), TPC-DS, and TPC-H. We ran experiments on a 40-node CloudLab cluster \cite{cloudlab}. We setup Tez atop YARN for the experiment. 
Furthermore, to understand performance at a larger scale, we used a trace-driven simulator to replay jobs from the same traces. Our key findings are:
\begin{itemize}[noitemsep,nolistsep]
  \item \name can closely approximate the {\burstq} performance of Strict Priority (\S\ref{sec:performane_guarantee}) and the long-term fairness for {\batchq}s of DRF (\S\ref{sec:fairness_guarantee}).

  \item \name handles multiple {\burstq}s to accommodate bounded priority and fairness (\S\ref{sec:admission}).

 \item \name can provide similar benefits in the large-scale setting (\S\ref{sec:performance_large_scale}).

 \item  When LQs have bursty arrivals of different sizes, BPF with the $\alpha$-strategy ensures the performance with the average usage remains similar (\S\ref{sec:performance_large_scale}).

 \item \new{Sensitivity analysis} shows that \name is robust to estimation errors (\S\ref{sec:estimation_error_sim}).

\end{itemize}

\subsection{Experimental Setup}

\begin{figure}[!ht]
	\centering	
	{\includegraphics[width=0.7\linewidth]{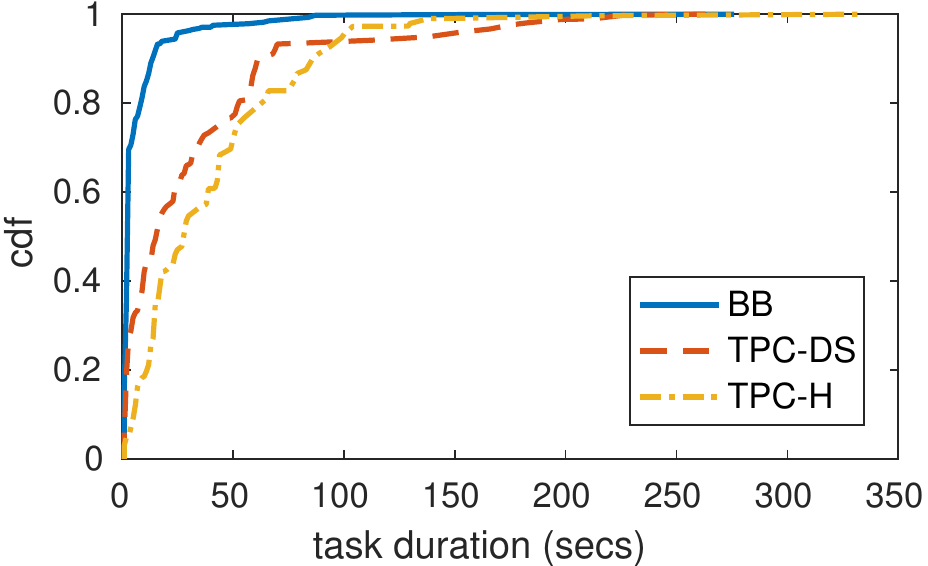}}
	\caption{CDFs of task durations across workloads.}
	\label{fig:worklad_cdf}
\end{figure}

\paragraph{Workloads} Our workloads consist of jobs from public benchmarks -- BigBench (BB) \cite{bbench}, TPC-DS \cite{tpc-ds}, and TPC-H \cite{tpc-h} traces.
A job has multiple stages.
A new stage can be executed if its prerequisite stages are finished.
A stage has a number of equivalent tasks in terms of resource demands and durations.
The cumulative distribution functions (CDFs) task durations across the three benchmarks are presented in Figure \ref{fig:worklad_cdf}.
In each experiment run, we chose the {\burstq} jobs from one of the traces such that their shortest completion times are less than 30 seconds.
We scale these jobs to make sure their instantaneous demands reach the maximum capacity of a single resource. The {\batchq} jobs are randomly chosen from one of the traces.
Each {\batchq} job lasts from tens of seconds to tens of minutes.
Each cluster experiment has 100 {\batchq} jobs, and each simulation experiment has 500 {\batchq} jobs.
Throughout the evaluation, all the {\batchq} jobs are submitted up at the beginning while the {\burstq} jobs arrive sequentially.
Our default experimental setup has a single {\burstq} and 8 {\batchq}s.

\textbf{User Input} Since the traces give us the resource demand and durations of the job tasks, we can set an ON period (i.e., when a {\burstq} job is active) equal to the shortest completion time of its corresponding {\burstq} job.
The average of ON periods is 27 seconds across the traces.
Without loss of generality, we assume that the {\burstq} jobs arrive periodically. 
The case of aperiodic {\burstq} jobs is similar to multiple {\burstq}s with different periods.
Unless otherwise noted, the inter-arrival period of two {\burstq} jobs is 300 seconds (1000 seconds) for the cluster experiment (the simulation experiment).

\textbf{Experimental Cluster}
We setup Apache Hadoop 2.7.2 (YARN) on a cluster having 40 worker nodes on CloudLab~\cite{cloudlab} (40-node cluster).
Each node has 32 CPU cores, 64 GB RAM, a 10 Gbps NIC, and runs Ubuntu 16.04.
Totally, the cluster has 1280 CPU cores and 2.5 TB memory.
The cluster also has a master node with the same specification running the resource manager (RM).


\textbf{Trace-driven Simulator} To have the experimental results on a larger scale, we build a simulator that mimics the system like Tez atop YARN.
The simulator can replay the directed acyclic graph jobs (like Tez does), and simulate the fair scheduler of YARN at queue level.
For the jobs in the same queue, we allocate the resource to them in a FIFO manner.
Unlike YARN, the simulator supports 6 resources, i.e., CPU, memory, disk in/out throughputs, and network in/out throughputs.

\textbf{Baselines} We compare \name against the following:
\begin{itemize}[noitemsep,nolistsep]
 \item \textbf{ Dominant Resource Fairness (DRF)}: DRF algorithm is implemented in YARN Fair Scheduler \cite{hadoop-fair-scheduler}.
DRF uses the concept of the dominant resource to compare multi-dimensional resources \cite{drf}.
The idea is that resource allocation should be determined by the dominant share of a queue, which is the maximum share of any resource (memory or CPU). Essentially, DRF seeks to maximize the minimum dominant share across all queues.

 \item  \textbf{Strict Priority (SP)}: We use Strict Priority to provide the best performance for {\burstq} jobs.
In fact, we borrow the concept of ``Strict Priority" from network traffic scheduling that enables Strict Priority queues to get bandwidth before other queues \cite{strict_priority}.
Similarly, we enable the {\burstq}s to receive all resources they need first, and then allocate the remaining resources to other queues. If there are conflicts among the {\burstq}s, we use DRF among them.

 \item  \textbf{Naive-\name (N-\name)}: N-\name is a simple version of \name that can provide bounded performance guarantee and fairness.
However, N-\name does not support admission to Soft Guarantee.
For the queues that satisfy the safety condition \ref{eqn:ad-safety}, N-\name decides to admit them to Hard Guarantee if they meet the fairness condition \ref{eqn:ad-fair} and resource condition \ref{eqn:ad-enough}.
Otherwise, it put the queues into the Elastic class. We use N-\name as a baseline when there are multiple {\burstq}s (\S\ref{sec:admission}).
\end{itemize}
Overall, SP is the upper bound in terms of performance guarantee, and DRF is the upper bound of fairness guarantee for our proposed approach.

\textbf{Metrics} Our primary metric is the \emph{average completion times} (avg. compl.) of {\burstq} jobs or {\batchq} jobs.
To show the performance improvement, we use the average completion times of {\burstq} jobs across the three approaches.
On the other hand, we use average completion times of {\batchq} jobs to show that \name also protects the {\batchq} jobs.
Additionally, we use \emph{factor of improvement} to show how much \name can speed up the {\burstq} jobs compared to DRF as
$$ \text{Factor of improvement} = \frac{\text{avg. compl. of DRF}}{\text{avg. compl. of \name}}. $$

\subsection{\name in Testbed Experiments}
\label{sec:testbed}

\subsubsection{\name in Practice}

\begin{figure}[!t]
    \centering
    
    \includegraphics[width=0.3\linewidth]{fig/b1_mov_legend} 
    \\
    \includegraphics[width=0.8\linewidth]{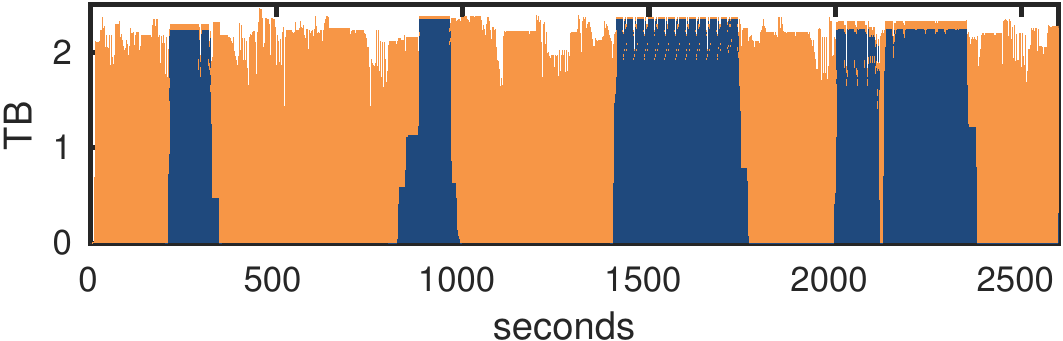} 
    
    \caption{[Cluster] \name's solution for the motivational problem (\S\ref{sec:ex}). The first two jobs of {\burstq} quickly finish and the last two jobs are prevented from using too much resource. This solution is close to the optimal one.}
     
    \label{fig:solution}
\end{figure}

Before diving into the details of our evaluation, recall the motivational problem from Section~\ref{sec:ex}. 
Figure~\ref{fig:solution} depicts how \name solves it in the testbed. \name enables the first two jobs of {\burstq} to quickly finish in 141 and 180 seconds.
For the two large jobs arriving at 1400 and 2000 seconds, the share is very large only in roughly 335 seconds but it is cut down to give back resource to {\batchq}.

\subsubsection{Performance Guarantee}
\label{sec:performane_guarantee}

\begin{figure}[!t]
\centering
\includegraphics[width=0.7\linewidth]{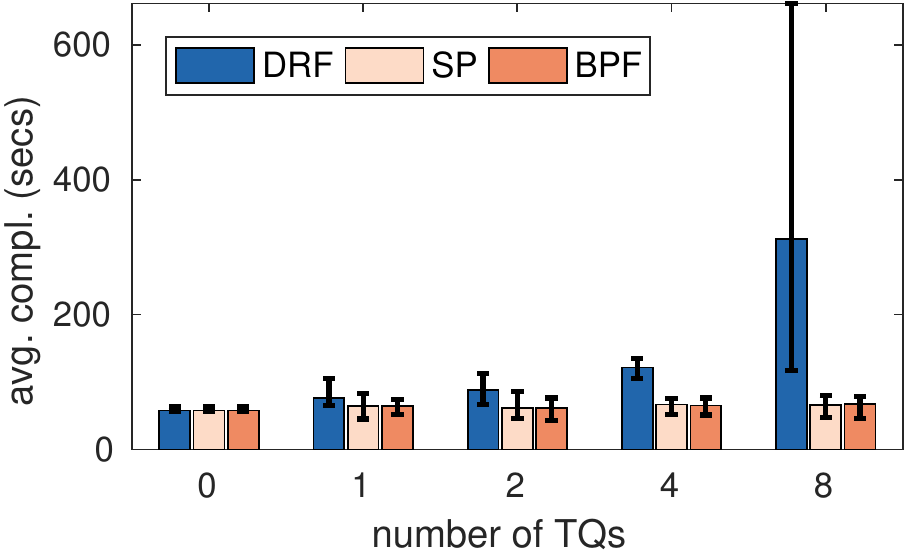}
\caption{[Cluster] Average completion time of \burstq jobs in a single {\burstq} across the 3 schedulers when varying the number of {\batchq}s. \name and SP guarantee the average completion time of the {\burstq} jobs while DRF significantly suffers from the increase of number of {\batchq}s.}
\label{fig:busty_perf_grt}
\end{figure}
 
Next, we focus on what happens when there are more than one {\batchq}.
Figure \ref{fig:busty_perf_grt} shows that average completion time of {\burstq} jobs in the 40-node cluster on the BB workload. In this setting, there are a single {\burstq} and multiple {\batchq}s. The x-axis shows the number of TQs in the cluster.

When there are no TQs, the average completion times of {\burstq} jobs across three schedulers are the same (57 seconds).
The completion times are greater than the average ON period (27 seconds) because of inefficient resource packing and allocation overheads.
In practice, the resource demand of tasks cannot utilize all the resources of a node that results in large unallocated resources across the cluster.
Hence, the {\burstq} jobs are not able to receive the whole cluster capacity as expected.
More importantly, this delay is also caused by allocation overheads, such as waiting for containers to be allocated or launching containers.

As the number of {\batchq}s increases, the performance of DRF significantly degrades because DRF tends to allocate less resource to {\burstq} jobs.
DRF is the worst among three schedulers. In contrast, \name and SP give the highest priority to {\burstq}s that guarantees the performance of {\burstq} jobs.
The average completion times, when TQs are available (1,2,4, and 8), are almost the same (65 seconds).
These average completion times are still larger than the case of no TQs because of non-preemption.
The {\burstq} jobs are not able to receive the resources that are still used by the running tasks. 

\begin{table}[!t]
\centering
\caption{[Cluster] Factor of improvement by \name across various workload with respect to the number of {\batchq}s.} 

\begin{tabular}{|c|c|c|c|c|} \hline
\small
\centering
Workload & 1 TQ  & 2 TQs & 4 TQs & 8 TQs \\ \hline \hline
BB & 1.18 & 1.42 & 1.86 & 4.66 \\ \hline 
TPC-DS &  1.35 &   1.61  &  2.29  &  5.38  \\ \hline 
TPC-H & 1.10  & 1.37 & 2.01 &  5.12\\ \hline 
\end{tabular}

\label{tbl:speed_up}
\end{table}

To understand how well \name performs on various workload traces, we carried out the same experiments on TPC-DS and TPC-H.
As SP and \name achieve similar performance, we only present the factors of improvement of \name across the various workloads in Table \ref{tbl:speed_up}.
The numbers on the table show consistent improvements inn terms of the average completion times of {\burstq} jobs. 

\begin{figure}[!t]
    \centering
    
    \subfloat[1 LQ \& 4 TQs]{\includegraphics[width=0.48\linewidth]{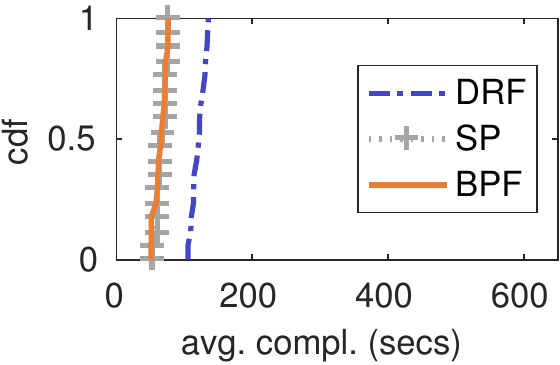} \label{fig:busty_perf_grt_cdf_a}}    
    \subfloat[1 LQ \& 8 TQs]{\includegraphics[width=0.48\linewidth]{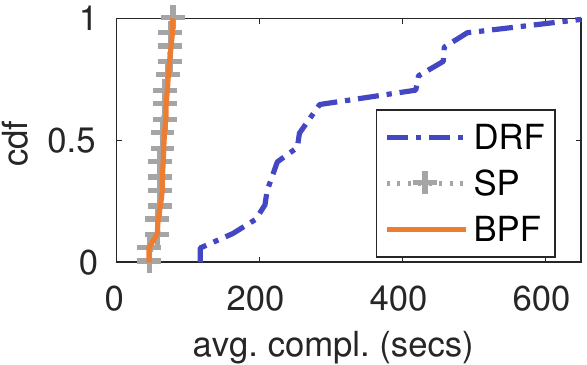} \label{fig:busty_perf_grt_cdf_b}}
    
    \caption{[Cluster] The completion time of LQ jobs is predictable using \name.}
    \label{fig:busty_perf_grt_cdf}
    
\end{figure}

In addition to the average completion time, we evaluated the performance of individual {\burstq} jobs.
Figure \ref{fig:busty_perf_grt_cdf} shows that cumulative distribution functions (cdf) of the completion times across 3 approaches.
Figure \ref{fig:busty_perf_grt_cdf_a} and \ref{fig:busty_perf_grt_cdf_b} are the experimental results for the cases of 4 {\batchq}s and 8 {\batchq}s, respectively.
We observe that the completion times of {\burstq} jobs in DRF are not stable and vary a lot when the number of {\burstq}s becomes large as in Figure \ref{fig:busty_perf_grt_cdf_b}.
The variation is caused by the instantaneous fairness and the variance of total resource demand.

\subsubsection{Fairness Guarantee}
\label{sec:fairness_guarantee}

\begin{figure}[!t]
	
    \centering
    \includegraphics[width=0.7\linewidth]{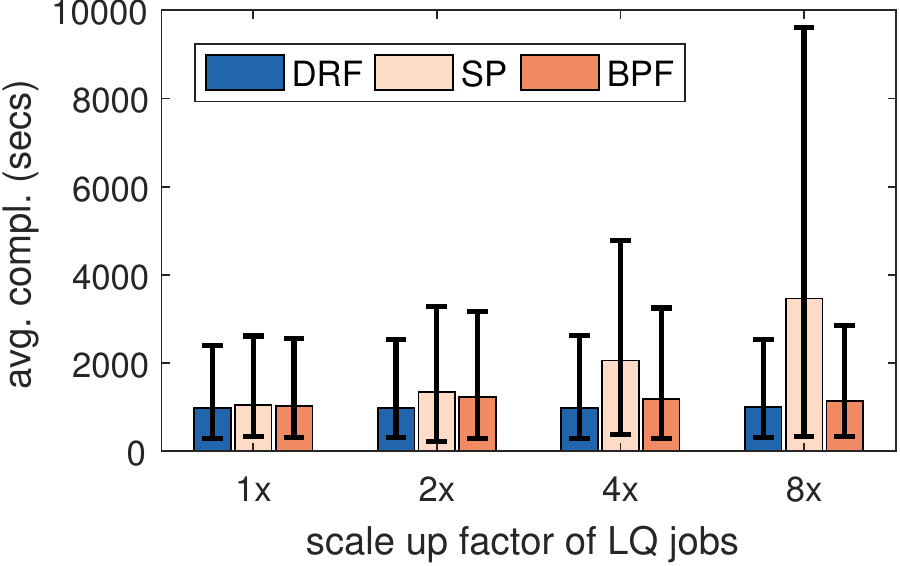} 

    \caption{[Cluster] \name protects the batch jobs up to $3.05\times$ compared to SP.}
     
    \label{fig:protecting_batch_jobs}
\end{figure}

Figure \ref{fig:protecting_batch_jobs} shows the average completion time of {\batchq} jobs when we scale up the number of tasks of {\burstq} jobs are by 1x, 2x, 4x, and 8x.
In this experiment, there are one {\burstq} and 8 {\batchq}s.

Since DRF is a fair scheduler, the average completion times of {\batchq} jobs are almost not affected by the size of {\burstq} jobs.
However, SP allocates too much resource to {\burstq} jobs that significantly hurts {\batchq} jobs.
Since SP provides the highest priority for the {\burstq} jobs, it makes the {\batchq} jobs to starve for resources.
\name performs closely to DRF.
While DRF maintains instantaneous fairness, \name maintains the long-term fairness among the queues.



\subsubsection{Scheduling Overheads}

Recall from Section \ref{sec:impl} that the \name scheduler has three components: user input, admission control, and allocation.
Compared to the default schedulers in YARN, our scheduler has additional scheduling overheads for admission control and additional computation in allocation.

Since we only implement our scheduler in the Resource Manager, the scheduling overheads occur at the master node.
To measure the scheduling overheads, we run admission control for 10000 {\burstq} queues and 10000 {\batchq} queues on a master node -- Intel Xeon E3 2.4 GHz (with 12 cores).
Each LQ queue has 500 ON/OFF cycles.
Recall the \textsc{LQAdmit} and \textsc{TQAdmit} functions in Algorithm \ref{algorithm1}, the admission overheads increase linearly to the number of queues.
The total admission overheads are approximately 1 ms, which is significantly smaller than the default update interval in YARN Fair Scheduler, i.e., 500 ms \cite{hadoop-fair-scheduler}.
The additional computation time spent in allocation is also negligible (less than 1 ms).


\subsubsection{Admission Control for Multiple LQs}
\label{sec:admission}

\begin{figure}[!h]
	\centering
	
	\includegraphics[width=0.6\linewidth]{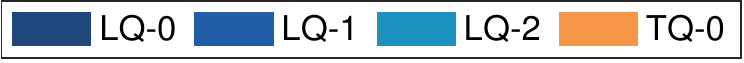} 
	
	\subfloat[DRF: \burstq-0, \burstq-1, \burstq-2 are unhappy with high latency.]{\includegraphics[width=0.8\linewidth]{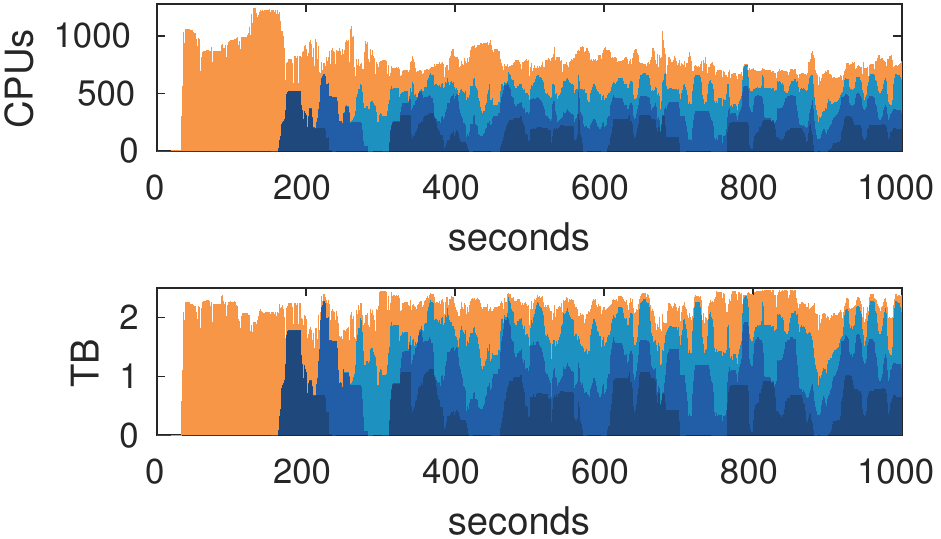} \label{fig:admission_drf_cluster}} 

	\subfloat[SP: \batchq-0 is starving of resources.]{\includegraphics[width=0.8\linewidth]{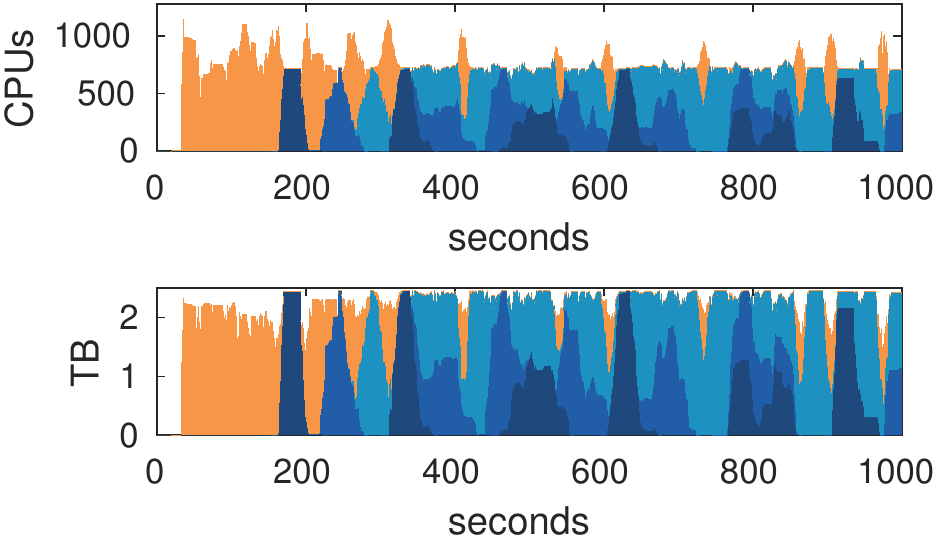} \label{fig:admission_strict_cluster}}
    
	\subfloat[N-\name: Only {\burstq}-0 and {\batchq}-0 are happy.]{\includegraphics[width=0.8\linewidth]{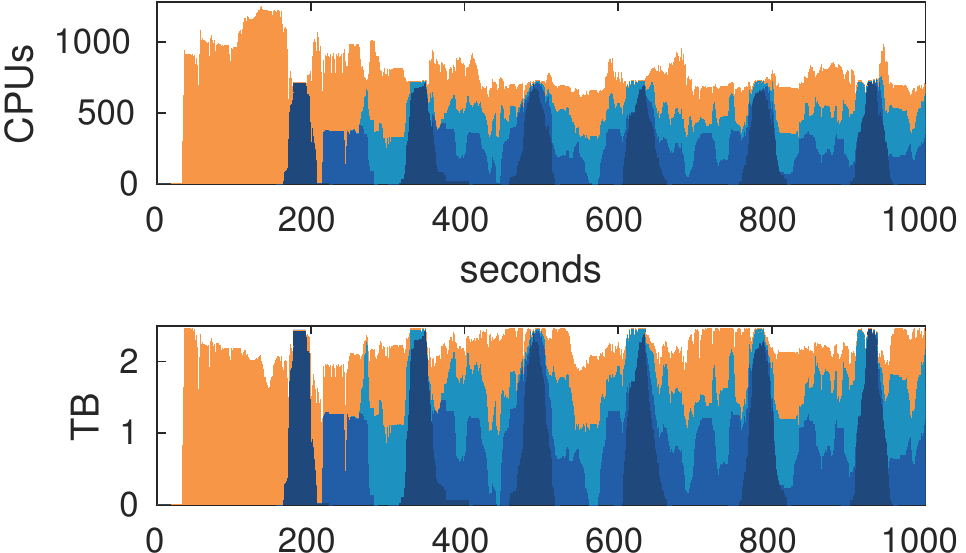} \label{fig:admission_hard_cluster}}
   
	\subfloat[\name: {\burstq}-0, {\burstq}-1 and {\batchq}-1 are happy.]{\includegraphics[width=0.8\linewidth]{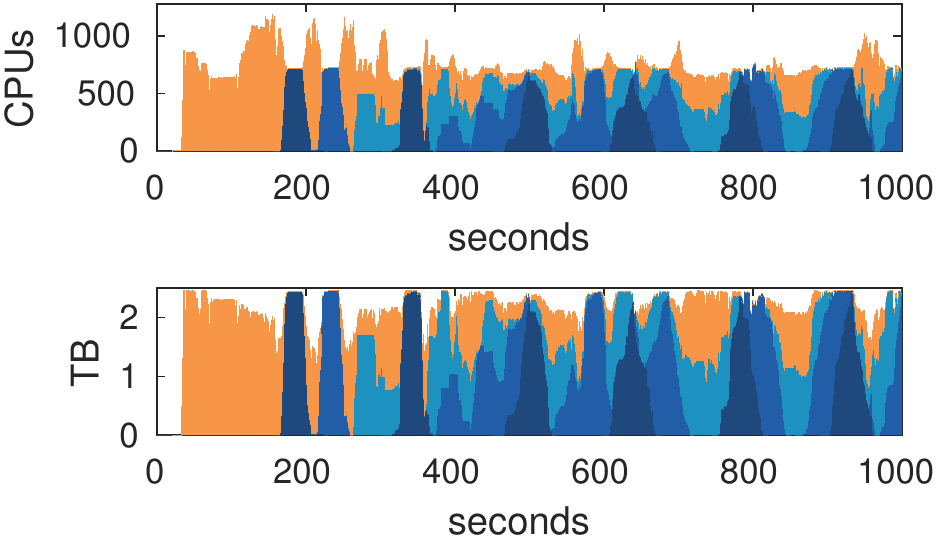} \label{fig:admission_speedfair_cluster}}
	
	\caption{[Cluster]. DRF and SP fail to guarantee both performance and fairness simultaneously. \name gives the best performance to \burstq-0, near optimal performance for \burstq-1, and maintains fairness among 4 queues. \burstq-2 requires too much resource, so its performance cannot be guaranteed.}
	 
	\label{fig:admission_control_cluster}
\end{figure}

To demonstrate how \name works with multiple {\burstq}s, we set up 3 {\burstq}s (\burstq-0, \burstq-1, and \burstq-2) and a single {\batchq} (\batchq-0).
The jobs \batchq-0 are queued up at the beginning while \burstq-0, \burstq-1, and \burstq-2 arrive at 50, 100, and 150 seconds, respectively.
The periods of {\burstq}-0, {\burstq}-1, and {\burstq}-2 are 150, 110, and 60 secs. All the {\burstq}s jobs have the identical demand and task durations.
The TQ jobs are chosen from the BB benchmark.
\name admits {\burstq}-0 to the Hard Guarantee class, {\burstq}-1 to the Soft Guarantee class, and {\burstq}-2 to the Elastic class.

Figure~\ref{fig:admission_drf_cluster} shows the resource usage for each queue across four schedulers, i.e., DRF, SP, N-\name and \name.
As an instantaneously fair scheduler, DRF continuously maintains the fair share for all queues as in Figure \ref{fig:admission_drf_cluster}.
Since \burstq-2 requires a lot of resources, SP makes \batchq-0 starving for resources (Figure \ref{fig:admission_strict_cluster}). N-\name provides \burstq-0 with resource guarantee and it fairly share the resources to \burstq-1, \burstq-2, and \batchq-0 (Figure \ref{fig:admission_hard_cluster}).
\name provides hard guarantee to \burstq-0 and soft guarantee to \burstq-1 as in Figure \ref{fig:admission_speedfair_cluster}.
The soft guarantee allows \burstq-1 to perform better than using N-\name.
Since \burstq-2 demands too much resources, \name treats it like \batchq-0.

\begin{figure}[!h]
	\centering
	
	\includegraphics[width=0.7\linewidth]{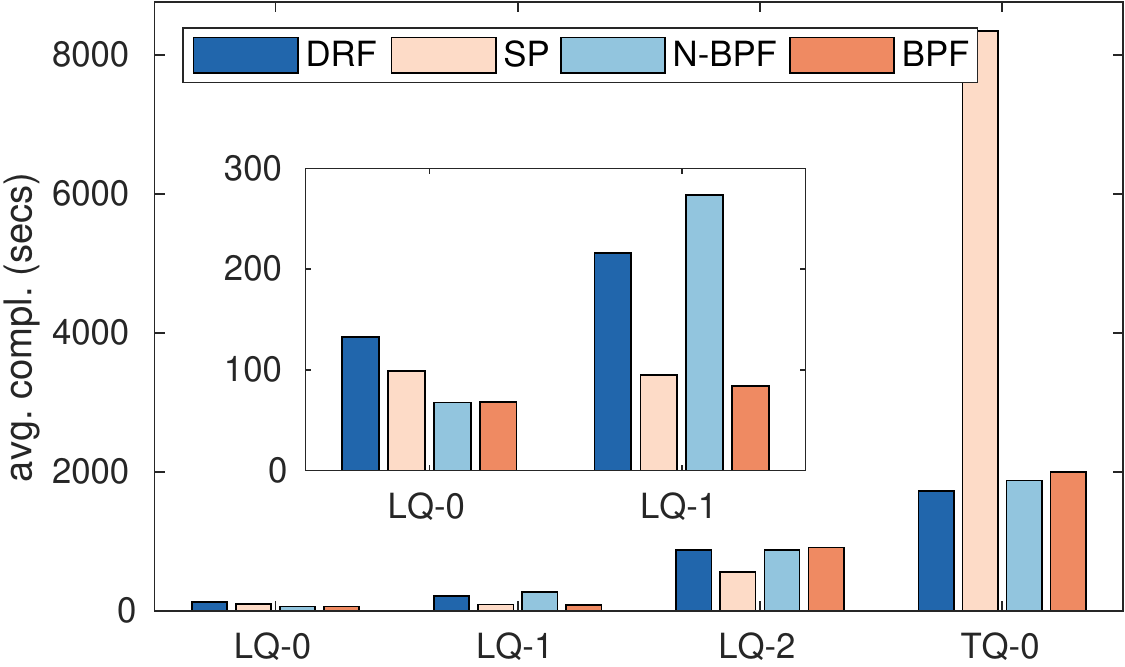}
	
	\caption{[Cluster] \name provides with better performance for {\burstq}s than DRF and N-\name. Unlike SP, \name protects the performance of {\batchq} jobs.}
	 
	\label{fig:avg_multi_queue_cluster}
\end{figure}

Figure \ref{fig:avg_multi_queue_cluster} shows the average completion time of jobs on each queue across the four schedulers.
The performance of DRF for {\burstq} jobs is the worst among the four schedulers but it is the best for only \batchq-0.
The performance of SP is good for {\burstq} jobs but it is the worst for {\batchq} jobs.
N-\name provides the best performance for \burstq-0 but not \burstq-1 and \burstq-2.
\name is the best among the four schedulers.
Three of the four queues, i.e., \burstq-0, \burstq-1, and \batchq-0, significantly benefit from \name.
\name even outperforms SP for \burstq-0 and \burstq-1 jobs and does not hurt {\batchq}.

\begin{figure*}[!t]
	\centering
	  
	\subfloat[Percentage of arrivals completed by the deadlines.]{\includegraphics[width=0.3\linewidth]{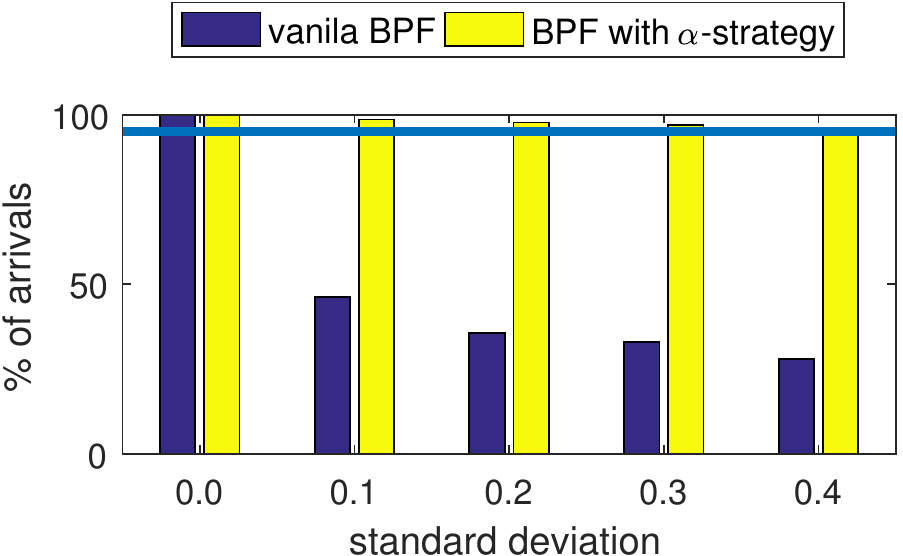} \label{fig:probabilistic_result}}
	\hspace{0.2cm}
	\subfloat[Requested demand normalized by that under the vanilla \name. ]{\includegraphics[width=0.3\linewidth]{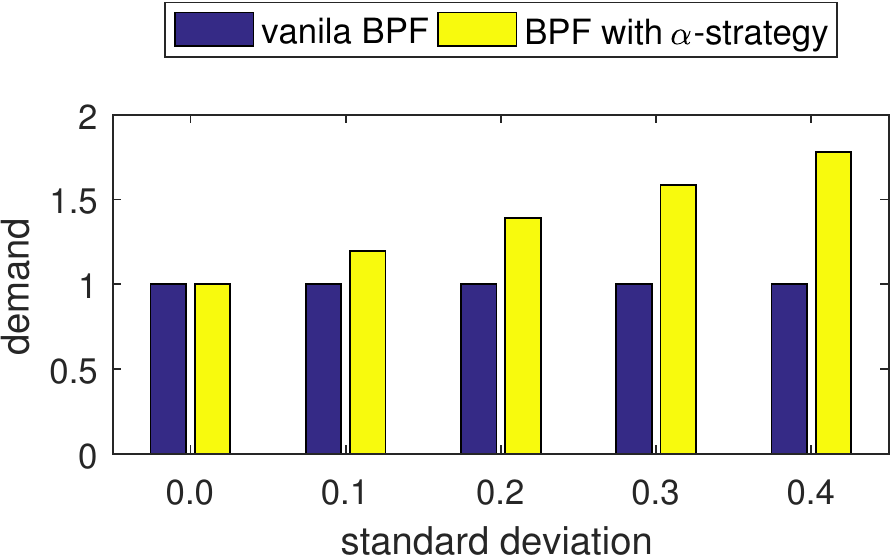} \label{fig:probabilistic_provision}}	
	\hspace{0.2cm}
	\subfloat[Resource consumption of LQ.]{\includegraphics[width=0.3\linewidth]{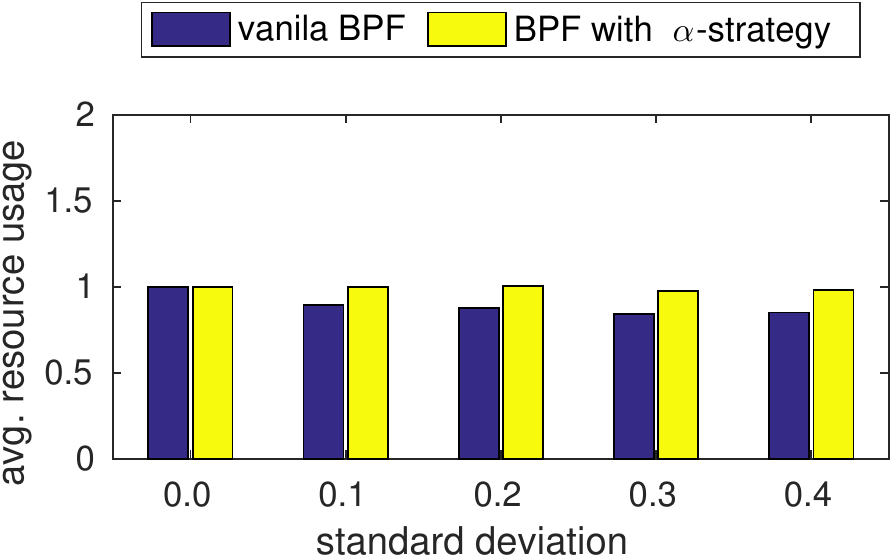} \label{fig:prob_avg_res}}
	
	\caption{[Simulation] The proposed $\alpha$-strategy under $\alpha$=95\% is robust against the uncertainties.}
	\label{fig:probabilistic}
	  
\end{figure*}

\subsection{\name in Trace-Driven Simulations}
\label{sec:performance_large_scale}


To verify the correctness of the large-scale simulator, we replayed the BB trace logs from cluster experiments in the simulator.
Table \ref{tbl:speed_up_sim} shows the factors of improvement in completion times of {\burstq} jobs from the simulator that are consistent with that from our cluster experiments (Table \ref{tbl:speed_up}). 

\begin{table}[!t]
\small
\centering
\begin{tabular}{|c|c|c|c|c|c|c|} \hline
\multirow{2}{*}{Workload} &  \multicolumn{5}{c}{Number of {\batchq}s} & \\ \hhline{~------}
 & 1 & 2 & 4 & 8 & 16 & 32 \\ \hline \hline
BB & 1.08  & 1.56 & 2.32 & 4.09 & 7.28 & 16.61  \\ \hline 
TPC-DS & 1.06 & 1.38 & 1.66 & 2.93 & 5.16 & 10.40 \\ \hline 
TPC-H  & 1.01 & 1.28 & 1.92 & 3.04 & 5.50 & 11.35 \\ \hline 
\end{tabular}
\caption{[Simulation] Factors of improvement by \name across various workloads w.r.t the number of {\batchq}s.} 
 
\label{tbl:speed_up_sim}
\end{table}

\name significantly improves over DRF when we have more {\batchq}s.
We note that the factors of improvement for TPC-DS and TPC-H in the simulation are smaller than that of the cluster experiments.
It turns out that DRF in TPC-DS and TPC-H suffers from allocation overheads that our simulation does not capture.
The allocation overheads for the {\burstq} jobs in TPC-DS and TPC-H are large because they have more stages than the {\burstq} jobs in BB (only 2 stages).


%

\todo{added estimation errors -- done.}
\subsubsection{Impact of Estimation Errors}
\label{sec:estimation_error_sim}

\name requires users to report their estimated demand for {\burstq} jobs.
\new{Demand estimation often results in estimation errors.}
To understand the impact of estimation errors on \name, we assume that estimation errors $e(\%)$ follow the standard normal distribution with zero mean.
The standard deviation (std.) of estimation errors lines in $[0, 50]$.
To adopt the estimation errors, we update the task demand and durations of {\burstq} jobs as $ {task}_{new} = {task}_{original}*(1+e/100)$.

\begin{figure}[!h]
	\centering
	\includegraphics[width=0.7\linewidth]{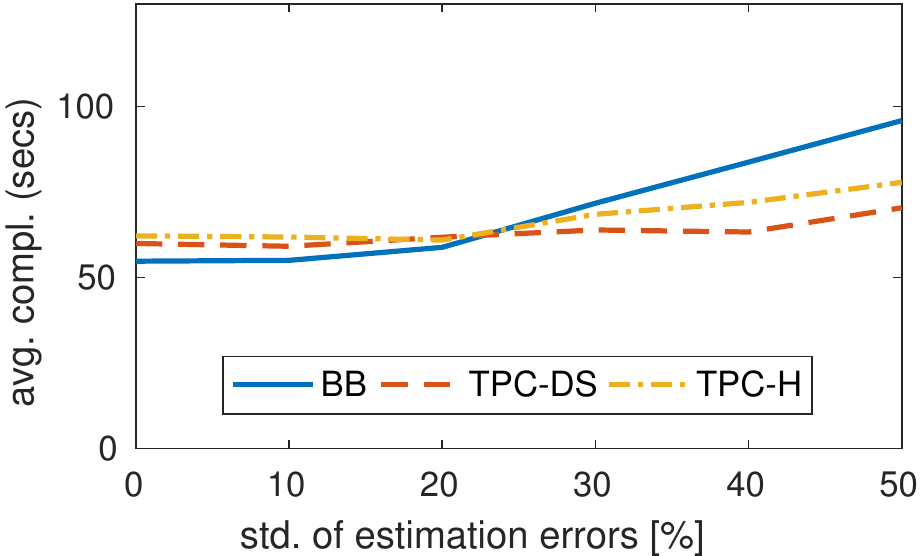}
	\caption{[Simulation] \name's performance degrades with larger estimation errors, yet is still significantly better than DRF (162 secs).}
	\label{fig:sen_analysis_est_err}
\end{figure}

Figure \ref{fig:sen_analysis_est_err} shows the impact of estimation errors on the average completion time of {\burstq} jobs.
There are 1 {\burstq} and 8 {\batchq}s.
{\burstq} jobs arrive every 350 seconds.
\name is robust when the standard deviation of estimation errors vary 0 to 20.
The {\burstq} jobs in BB suffer more from the large estimation errors (std. $>30$) than that of TPC-DS and TPC-H.
The delays are caused by the underestimated jobs because the excessive demand is not guaranteed by the system.
Meanwhile, the overestimated jobs do not suffer any delays as the guaranteed resource is more than needed.
Although estimation errors result in performance degradation, the performance of {\burstq} jobs is still much better than that of DRF (162 seconds).

\subsubsection{Performance of the $\alpha$-Strategy.}

Figure~\ref{fig:probabilistic} depicts the requested demand, performance, and resource usage under the vanilla \name and the one with the $\alpha$-strategy when arrivals have different sizes. In particular, as the variance increases, the vanilla \name can no longer complete $\alpha$ arrivals before the deadline. Actually, even with 10\% standard deviation, the percentage drops below 50\%. On the other side, \name with $\alpha$-strategy always satisfy the $\alpha$ requirement. Even though the reported demand increases, the average resource usage does not change much, e.g., {\batchq} receives the same long-term share.



\section{Related Work}
\label{sec:related}



\textbf{Bursty Applications in Big Data Clusters}
Big data clusters experience burstiness from a variety of sources, including periodic jobs \cite{jockey, rope, scarlett, omega}, interactive user sessions \cite{splunk-analysis}, as well as streaming applications \cite{spark-streaming, millwheel, trident}. 
Some of them show predictability in terms of inter-arrival times between successive jobs (\eg, Spark Streaming \cite{spark-streaming} runs periodic mini batches in regular intervals), while some others follow different arrival processes (\eg, user interactions with large datasets \cite{splunk-analysis}).
Similarly, resource requirements of the successive jobs can sometimes be predictable, but often it can be difficult to predict due to external load variations (\eg, time-of-day or similar patterns); the latter, without {\name}, can inadvertently hurt batch queues (\S\ref{sec:motivation}).

\textbf{Multi-Resource Job Schedulers}
Although early jobs schedulers dealt with a single resource \cite{late, mantri, quincy}, modern cluster resource managers, \eg, Mesos \cite{mesos}, YARN \cite{yarn}, and others \cite{omega, borg, cosmos}, employ multi-resource schedulers \cite{drf, sjf, tetris, carbyne, apollo, mercury, hdrf} to handle multiple resources and optimize diverse objectives. 
These objectives can be fairness (\eg, DRF \cite{drf}), performance (\eg, shortest-job-first (SJF) \cite{sjf}), efficiency (\eg, Tetris \cite{tetris}), or different combinations of the three (\eg, Carbyne \cite{carbyne}).
\new{Hawk \cite{hawk} focuses on reducing the overheads in scheduling the large number of small jobs.
Chen \textit{et al.} design a preemption algorithm to prioritize short jobs without resource guarantee \cite{chen2017preemptive}.}
However, \emph{all} of these focus on instantaneous objectives, with instantaneous fairness being the most common goal. 
To the best of our knowledge, {\name} is the first multi-resource job scheduler with long-term memory.

\textbf{Handling Burstiness}
Supporting multiple classes of traffic is a classic networking problem that, over the years, have arisen in local area networks \cite{cbq, intserv-hierarchy, hfsc, diffserv-rfc2475, wfq}, wide area networks \cite{bwe, b4, swan}, and in datacenter networks \cite{silo, qjump}. 
All of them employ some form of admission control to provide quality-of-service guarantees.
They consider only a single link (\ie, a single resource). 
In contrast, {\name} considers multi-resource jobs and builds on top this large body of literature.

BVT \cite{bvt} is\new{ a thread-based CPU scheduler} that was designed to work with both real-time and best-effort tasks. Although it prioritizes the real-time tasks, it cannot guarantee performance and fairness.

\textbf{Collocating Mixed Workloads in Datacenters}
\new{
Latency-sensitive and best-effort workloads are often collocated. Heracles \cite{heracles} and Parties \cite{parties} handle mixed workloads to increase the utilization of servers. Bistro \cite{bistro} allows both data-intensive and online workloads to share the same infrastructure. Morpheus \cite{morpheus} reserves resources for periodic jobs ahead of time. All of them prioritize the latency-sensitive workloads to meet the quality of service requirement but do not provide both resource guarantee and fairness.}

\textbf{Expressing Burstiness Requirements}
{\name} is not the first system that allows users to express their time-varying resource requirements. 
Similar challenges have appeared in traditional networks \cite{hfsc}, network calculus \cite{cruz1, cruz2}, datacenters \cite{silo, pulsar}, and wide-area networks \cite{bwe}.
Akin to them, {\name} requires users to explicitly provide their burst durations and sizes; {\name} tries to enforce those requirements in short and long terms. 
Unlike them, however, {\name} explores how to allow users to express their requirements in a multi-dimensional space, where each dimension corresponds to individual resources. 
One possible way to collapse the multi-dimensional interface to a single dimension is using the \emph{progress} \cite{hug, drf}; however, progress only applies to scenarios when a user's utility is captured using Leontief preferences.

\section{Conclusion}
\label{sec:outro}

To enable the coexist of latency-sensitive {\burstq}s and the {\batchq}s, we proposed \name (Bounded Priority Fairness).
\name provides bounded performance guarantee to {\burstq}s and maintains the long-term fairness for {\batchq}s.
\name classifies the queues into three classes: Hard Guarantee, Soft Guarantee and Elastic.
\name provides the best performance to {\burstq}s in the Hard Guarantee class and the better performance for {\burstq}s in the Soft Guarantee class. The scheduling is executed in a strategyproof manner, which is critical for public clouds. 
The queues in the Elastic class share the left-over resources to maximize the system utilization. 
In the deployments, we show that \name not only outperforms the DRF up to $5.38\times$ for {\burstq} jobs but also protects {\batchq} jobs up to $3.05\times$ compared to Strict Priority. When {\burstq}'s arrivals have different sizes, adding the $\alpha$-strategy can satisfy the deadlines with similar resource utilization.


\phantomsection
\label{EndOfPaper}









\section{Acknowledgments}
This research is supported by NSF grants CNS-1617773, 1730128, 1919752, 1717588, 1617698 and was partially funded by MSIT, Korea, IITP-2019-2011-1-00783.
\section{Appendices}
\subsection{Proof of strategyproofness}
\label{proof:strategyproof}

Let the true demand and deadline of LQ-$i$ be $(\myvec{d},t)$ for a particular arrival. Let the request parameter be $(\myvec{v},t')$. We first argue that $\myvec{v}=\beta \myvec{d}$ holds with $\beta>0$.

As $\myvec{d}=\langle d^1,d^2,\cdots,d^k \rangle,\myvec{v}=\langle v^1,v^2,\cdots,v^k \rangle$, let $p= \min_k \left\{\frac{v^i}{d^i}\right\}$. Define a new vector $\myvec{z}= \langle z^1,z^2,\cdots,z^k \rangle$ and let $z^i  = pd^i$. Notice here $\myvec{z}$ has the same performance as $\myvec{v}$, while $z_i \leq v_i$ for any $i$. Therefore, $\myvec{z}$ may request no more resources than $\myvec{v}$, which is more likely to be admitted. Hence, it is always better to request $\myvec{z}$, which is proportional to $\myvec{d}$, the true demand.

Regarding the demand $\myvec{d}$, reporting a larger $\myvec{v}$ ($\beta>1$) still satisfies its demand, while has a higher risk being rejected as it requests higher demand, while it does not make sense to report a smaller $\myvec{v}$ ($\beta<1$) as it may receive fewer resources than it actually needs.
Therefore, there is no incentive to for LQ-$i$ to lie about its $\myvec{d}$.
The argument for deadline is similar. Reporting a larger deadline does not make sense as it may receive fewer resources than it actually needs. On the other hand, reporting a tighter deadline still satisfies the deadlines, while has a higher risk being rejected as it requests higher demand. 
Therefore, there is no incentive to for LQ-$i$ to lie about its deadline, either.

\label{EndOfPaper}

\bibliographystyle{abbrv}
\bibliography{bib/refs}

\begin{thebibliography}{10}

\bibitem{hadoop}
{{Apache Hadoop}}.
\newblock \url{http://hadoop.apache.org}.

\bibitem{hadoop-fair-scheduler}
{YARN Fair Scheduler}.
\newblock
  \url{http://hadoop.apache.org/docs/r2.4.1/hadoop-yarn/hadoop-yarn-site/FairScheduler.html},
  2014.

\bibitem{bbench}
{Big-Data-Benchmark-for-Big-Bench}.
\newblock
  \url{https://github.com/intel-hadoop/Big-Data-Benchmark-for-Big-Bench}, 2016.

\bibitem{presto}
{Presto: Distributed {SQL} Query Engine for Big Data}.
\newblock \url{https://prestodb.io/}, 2016.

\bibitem{tez}
{{Apache Tez}}.
\newblock \url{http://tez.apache.org}, 2017.

\bibitem{cloudlab}
{{Cloudlab}}.
\newblock \url{http://www.cloudlab.us/}, 2017.

\bibitem{tpc-ds}
{TPC Benchmark DS (TPC-DS)}.
\newblock \url{http://www.tpc.org/tpcds}, 2017.

\bibitem{tpc-h}
{TPC Benchmark H (TPC-H)}.
\newblock \url{http://www.tpc.org/tpch}, 2017.

\bibitem{trident}
Trident: Stateful stream processing on {Storm}.
\newblock \url{http://storm.apache.org/documentation/Trident-tutorial.html},
  2017.

\bibitem{rope}
S.~Agarwal, S.~Kandula, N.~Burno, M.-C. Wu, I.~Stoica, and J.~Zhou.
\newblock Re-optimizing data parallel computing.
\newblock In {\em NSDI}, 2012.

\bibitem{blinkdb}
S.~Agarwal, B.~Mozafari, A.~Panda, H.~Milner, S.~Madden, and I.~Stoica.
\newblock {BlinkDB}: Queries with bounded errors and bounded response times on
  very large data.
\newblock In {\em EuroSys}, 2013.

\bibitem{millwheel}
T.~Akidau, A.~Balikov, K.~Bekiro{\u{g}}lu, S.~Chernyak, J.~Haberman, R.~Lax,
  S.~McVeety, D.~Mills, P.~Nordstrom, and S.~Whittle.
\newblock {MillWheel}: Fault-tolerant stream processing at {Internet} scale.
\newblock 2013.

\bibitem{cherrypick}
O.~Alipourfard, H.~H. Liu, J.~Chen, S.~Venkataraman, M.~Yu, and M.~Zhang.
\newblock Cherrypick: Adaptively unearthing the best cloud configurations for
  big data analytics.
\newblock In {\em 14th $\{$USENIX$\}$ Symposium on Networked Systems Design and
  Implementation ($\{$NSDI$\}$ 17)}, pages 469--482, 2017.

\bibitem{splunk-analysis}
S.~Alspaugh, B.~Chen, J.~Lin, A.~Ganapathi, M.~Hearst, and R.~Katz.
\newblock Analyzing log analysis: An empirical study of user log mining.
\newblock In {\em LISA}, 2014.

\bibitem{scarlett}
G.~Ananthanarayanan, S.~Agarwal, S.~Kandula, A.~Greenberg, I.~Stoica,
  D.~Harlan, and E.~Harris.
\newblock Scarlett: Coping with skewed popularity content in {MapReduce}
  clusters.
\newblock In {\em EuroSys}, 2011.

\bibitem{pacman}
G.~Ananthanarayanan, A.~Ghodsi, A.~Wang, D.~Borthakur, S.~Kandula, S.~Shenker,
  and I.~Stoica.
\newblock {PACMan}: Coordinated memory caching for parallel jobs.
\newblock In {\em NSDI}, 2012.

\bibitem{mantri}
G.~Ananthanarayanan, S.~Kandula, A.~Greenberg, I.~Stoica, Y.~Lu, B.~Saha, and
  E.~Harris.
\newblock Reining in the outliers in {MapReduce} clusters using {Mantri}.
\newblock In {\em OSDI}, 2010.

\bibitem{pulsar}
S.~Angel, H.~Ballani, T.~Karagiannis, G.~O'Shea, and E.~Thereska.
\newblock End-to-end performance isolation through virtual datacenters.
\newblock In {\em OSDI}, 2014.

\bibitem{bansal2001analysis}
N.~Bansal and M.~Harchol-Balter.
\newblock {\em Analysis of SRPT scheduling: Investigating unfairness},
  volume~29.
\newblock ACM, 2001.

\bibitem{hdrf}
A.~A. Bhattacharya, D.~Culler, E.~Friedman, A.~Ghodsi, S.~Shenker, and
  I.~Stoica.
\newblock Hierarchical scheduling for diverse datacenter workloads.
\newblock In {\em SoCC}, 2013.

\bibitem{diffserv-rfc2475}
S.~Blake, D.~Black, M.~Carlson, E.~Davies, Z.~Wang, and W.~Weiss.
\newblock {An Architecture for Differentiated Services}.
\newblock RFC 2475 (Informational), Dec. 1998.
\newblock Updated by RFC 3260.

\bibitem{apollo}
E.~Boutin, J.~Ekanayake, W.~Lin, B.~Shi, J.~Zhou, Z.~Qian, M.~Wu, and L.~Zhou.
\newblock Apollo: Scalable and coordinated scheduling for cloud-scale
  computing.
\newblock In {\em OSDI}, 2014.

\bibitem{flink}
P.~Carbone, S.~Ewen, S.~Haridi, A.~Katsifodimos, V.~Markl, and K.~Tzoumas.
\newblock {Apache Flink}: Stream and batch processing in a single engine.
\newblock {\em Data Engineering}, 2015.

\bibitem{cosmos}
R.~Chaiken, B.~Jenkins, P.~Larson, B.~Ramsey, D.~Shakib, S.~Weaver, and
  J.~Zhou.
\newblock {SCOPE}: Easy and efficient parallel processing of massive datasets.
\newblock In {\em VLDB}, 2008.

\bibitem{chaudhuri2004estimating}
S.~Chaudhuri, V.~Narasayya, and R.~Ramamurthy.
\newblock Estimating progress of execution for sql queries.
\newblock In {\em Proceedings of the 2004 ACM SIGMOD international conference
  on Management of data}, pages 803--814. ACM, 2004.

\bibitem{parties}
S.~Chen, C.~Delimitrou, and J.~F. Mart{\'\i}nez.
\newblock Parties: Qos-aware resource partitioning for multiple interactive
  services.
\newblock 2019.

\bibitem{chen2017preemptive}
W.~Chen, J.~Rao, and X.~Zhou.
\newblock Preemptive, low latency datacenter scheduling via lightweight
  virtualization.
\newblock In {\em 2017 $\{$USENIX$\}$ Annual Technical Conference
  ($\{$USENIX$\}$$\{$ATC$\}$ 17)}, pages 251--263, 2017.

\bibitem{hug}
M.~Chowdhury, Z.~Liu, A.~Ghodsi, and I.~Stoica.
\newblock {HUG}: Multi-resource fairness for correlated and elastic demands.
\newblock In {\em NSDI}, 2016.

\bibitem{aalo}
M.~Chowdhury and I.~Stoica.
\newblock Efficient coflow scheduling without prior knowledge.
\newblock In {\em SIGCOMM}, 2015.

\bibitem{orchestra}
M.~Chowdhury, M.~Zaharia, J.~Ma, M.~I. Jordan, and I.~Stoica.
\newblock Managing data transfers in computer clusters with {Orchestra}.
\newblock In {\em SIGCOMM}, 2011.

\bibitem{cruz1}
R.~Cruz.
\newblock A calculus for network delay, {Part I}: Network elements in
  isolation.
\newblock {\em IEEE Transactions on Information Theory}, 37(1):114--131, 1991.

\bibitem{cruz2}
R.~Cruz.
\newblock A calculus for network delay, {Part II}: Network analysis.
\newblock {\em IEEE Transactions on Information Theory}, 37(1):132--141, 1991.

\bibitem{mapreduce}
J.~Dean and S.~Ghemawat.
\newblock {MapReduce}: Simplified data processing on large clusters.
\newblock In {\em OSDI}, 2004.

\bibitem{hawk}
P.~Delgado, F.~Dinu, A.-M. Kermarrec, and W.~Zwaenepoel.
\newblock Hawk: Hybrid datacenter scheduling.
\newblock In {\em 2015 $\{$USENIX$\}$ Annual Technical Conference
  ($\{$USENIX$\}$$\{$ATC$\}$ 15)}, pages 499--510, 2015.

\bibitem{wfq}
A.~Demers, S.~Keshav, and S.~Shenker.
\newblock Analysis and simulation of a fair queueing algorithm.
\newblock In {\em SIGCOMM}, 1989.

\bibitem{bvt}
K.~J. Duda and D.~R. Cheriton.
\newblock Borrowed-virtual-time ({BVT}) scheduling: supporting
  latency-sensitive threads in a general-purpose scheduler.
\newblock {\em ACM SIGOPS Operating Systems Review}, 33(5):261--276, 1999.

\bibitem{jockey}
A.~D. Ferguson, P.~Bodik, S.~Kandula, E.~Boutin, and R.~Fonseca.
\newblock Jockey: Guaranteed job latency in data parallel clusters.
\newblock In {\em Eurosys}, 2012.

\bibitem{cbq}
S.~Floyd and V.~Jacobson.
\newblock Link-sharing and resource management models for packet networks.
\newblock {\em IEEE/ACM Transactions on Networking}, 3(4):365--386, 1995.

\bibitem{sjf}
M.~R. Garey, D.~S. Johnson, and R.~Sethi.
\newblock The complexity of flowshop and jobshop scheduling.
\newblock {\em Mathematics of Operations Research}, 1(2):117--129, 1976.

\bibitem{drfq}
A.~Ghodsi, V.~Sekar, M.~Zaharia, and I.~Stoica.
\newblock Multi-resource fair queueing for packet processing.
\newblock {\em SIGCOMM}, 2012.

\bibitem{drf}
A.~Ghodsi, M.~Zaharia, B.~Hindman, A.~Konwinski, S.~Shenker, and I.~Stoica.
\newblock Dominant resource fairness: Fair allocation of multiple resource
  types.
\newblock In {\em NSDI}, 2011.

\bibitem{choosy}
A.~Ghodsi, M.~Zaharia, S.~Shenker, and I.~Stoica.
\newblock Choosy: Max-min fair sharing for datacenter jobs with constraints.
\newblock In {\em EuroSys}, 2013.

\bibitem{bistro}
A.~Goder, A.~Spiridonov, and Y.~Wang.
\newblock Bistro: Scheduling data-parallel jobs against live production
  systems.
\newblock In {\em 2015 $\{$USENIX$\}$ Annual Technical Conference
  ($\{$USENIX$\}$$\{$ATC$\}$ 15)}, pages 459--471, 2015.

\bibitem{tetris}
R.~Grandl, G.~Ananthanarayanan, S.~Kandula, S.~Rao, and A.~Akella.
\newblock Multi-resource packing for cluster schedulers.
\newblock In {\em SIGCOMM}, 2014.

\bibitem{carbyne}
R.~Grandl, M.~Chowdhury, A.~Akella, and G.~Ananthanarayanan.
\newblock Altruistic scheduling in multi-resource clusters.
\newblock In {\em OSDI}, 2016.

\bibitem{graphene}
R.~Grandl, S.~Kandula, S.~Rao, A.~Akella, and J.~Kulkarni.
\newblock Graphene: Packing and dependency-aware scheduling for data-parallel
  clusters.
\newblock In {\em OSDI}, 2016.

\bibitem{qjump}
M.~P. Grosvenor, M.~Schwarzkopf, I.~Gog, R.~N. Watson, A.~W. Moore, S.~Hand,
  and J.~Crowcroft.
\newblock Queues don't matter when you can {JUMP} them!
\newblock In {\em NSDI}, 2015.

\bibitem{mesos}
B.~Hindman, A.~Konwinski, M.~Zaharia, A.~Ghodsi, A.~D. Joseph, R.~Katz,
  S.~Shenker, and I.~Stoica.
\newblock {Mesos: A Platform for Fine-Grained Resource Sharing in the Data
  Center}.
\newblock In {\em NSDI}, 2011.

\bibitem{swan}
C.-Y. Hong, S.~Kandula, R.~Mahajan, M.~Zhang, V.~Gill, M.~Nanduri, and
  R.~Wattenhofer.
\newblock Achieving high utilization with software-driven {WAN}.
\newblock In {\em SIGCOMM}, 2013.

\bibitem{dryad}
M.~Isard, M.~Budiu, Y.~Yu, A.~Birrell, and D.~Fetterly.
\newblock Dryad: Distributed data-parallel programs from sequential building
  blocks.
\newblock In {\em EuroSys}, 2007.

\bibitem{quincy}
M.~Isard, V.~Prabhakaran, J.~Currey, U.~Wieder, K.~Talwar, and A.~Goldberg.
\newblock Quincy: Fair scheduling for distributed computing clusters.
\newblock In {\em SOSP}, 2009.

\bibitem{jaffe-maxmin}
J.~M. Jaffe.
\newblock Bottleneck flow control.
\newblock {\em IEEE Transactions on Communications}, 29(7):954--962, 1981.

\bibitem{jaillet2012online}
P.~Jaillet and M.~R. Wagner.
\newblock {\em Online Optimization}.
\newblock Springer Publishing Company, Incorporated, 2012.

\bibitem{b4}
S.~Jain, A.~Kumar, S.~Mandal, J.~Ong, L.~Poutievski, A.~Singh, S.~Venkata,
  J.~Wanderer, J.~Zhou, M.~Zhu, et~al.
\newblock B4: Experience with a globally-deployed software defined {WAN}.
\newblock In {\em SIGCOMM}, 2013.

\bibitem{silo}
K.~Jang, J.~Sherry, H.~Ballani, and T.~Moncaster.
\newblock Silo: Predictable message completion time in the cloud.
\newblock In {\em SIGCOMM}, 2015.

\bibitem{multiresource-mungchiang}
C.~Joe-Wong, S.~Sen, T.~Lan, and M.~Chiang.
\newblock Multi-resource allocation: Fairness-efficiency tradeoffs in a
  unifying framework.
\newblock In {\em INFOCOM}, 2012.

\bibitem{morpheus}
S.~A. Jyothi, C.~Curino, I.~Menache, S.~M. Narayanamurthy, A.~Tumanov,
  J.~Yaniv, R.~Mavlyutov, {\'I}.~Goiri, S.~Krishnan, J.~Kulkarni, et~al.
\newblock Morpheus: Towards automated slos for enterprise clusters.
\newblock In {\em 12th $\{$USENIX$\}$ Symposium on Operating Systems Design and
  Implementation ($\{$OSDI$\}$ 16)}, pages 117--134, 2016.

\bibitem{mercury}
K.~Karanasos, S.~Rao, C.~Curino, C.~Douglas, K.~Chaliparambil, G.~Fumarola,
  S.~Heddaya, R.~Ramakrishnan, and S.~Sakalanaga.
\newblock Mercury: Hybrid centralized and distributed scheduling in large
  shared clusters.
\newblock In {\em USENIX ATC}, 2015.

\bibitem{strict_priority}
L.~Kleinrock and R.~Gail.
\newblock {\em Queueing systems: Problems and Solutions}.
\newblock Wiley, 1996.

\bibitem{konig2011statistical}
A.~C. K{\"o}nig, B.~Ding, S.~Chaudhuri, and V.~Narasayya.
\newblock A statistical approach towards robust progress estimation.
\newblock {\em Proceedings of the VLDB Endowment}, 5(4):382--393, 2011.

\bibitem{bwe}
A.~Kumar, S.~Jain, U.~Naik, A.~Raghuraman, N.~Kasinadhuni, E.~C. Zermeno, C.~S.
  Gunn, J.~Ai, B.~Carlin, M.~Amarandei-Stavila, et~al.
\newblock {BwE}: Flexible, hierarchical bandwidth allocation for {WAN}
  distributed computing.
\newblock In {\em SIGCOMM}, 2015.

\bibitem{heracles}
D.~Lo, L.~Cheng, R.~Govindaraju, P.~Ranganathan, and C.~Kozyrakis.
\newblock Heracles: improving resource efficiency at scale.
\newblock In {\em ISCA}, 2015.

\bibitem{graphlab}
Y.~Low, J.~Gonzalez, A.~Kyrola, D.~Bickson, C.~Guestrin, and J.~M. Hellerstein.
\newblock {GraphLab}: A new framework for parallel machine learning.
\newblock In {\em UAI}, 2010.

\bibitem{luo2004toward}
G.~Luo, J.~F. Naughton, C.~J. Ellmann, and M.~W. Watzke.
\newblock Toward a progress indicator for database queries.
\newblock In {\em Proceedings of the 2004 ACM SIGMOD international conference
  on Management of data}, pages 791--802. ACM, 2004.

\bibitem{paratimer}
K.~Morton, M.~Balazinska, and D.~Grossman.
\newblock {ParaTimer}: A progress indicator for {MapReduce} {DAGs}.
\newblock In {\em SIGMOD}, 2010.

\bibitem{moulin2014cooperative}
H.~Moulin.
\newblock {\em Cooperative microeconomics: a game-theoretic introduction}.
\newblock Princeton University Press, 2014.

\bibitem{naiad}
D.~G. Murray, F.~McSherry, R.~Isaacs, M.~Isard, P.~Barham, and M.~Abadi.
\newblock Naiad: A timely dataflow system.
\newblock In {\em SOSP}, 2013.

\bibitem{schneider2007stochastic}
J.~Schneider and S.~Kirkpatrick.
\newblock {\em Stochastic Optimization}.
\newblock Springer Science \& Business Media, 2007.

\bibitem{omega}
M.~Schwarzkopf, A.~Konwinski, M.~Abd-El-Malek, and J.~Wilkes.
\newblock Omega: Flexible, scalable schedulers for large compute clusters.
\newblock In {\em EuroSys}, 2013.

\bibitem{intserv-hierarchy}
S.~Shenker, D.~D. Clark, and L.~Zhang.
\newblock A scheduling service model and a scheduling architecture for an
  integrated services packet network.
\newblock Technical report, Xerox PARC, 1993.

\bibitem{hfsc}
I.~Stoica, H.~Zhang, and T.~S.~E. Ng.
\newblock {A} hierarchical fair service curve algorithm for link-sharing,
  real-time and priority service.
\newblock In {\em {SIGCOMM}}, 1997.

\bibitem{yarn}
V.~K. Vavilapalli, A.~C. Murthy, C.~Douglas, S.~Agarwal, M.~Konar, R.~Evans,
  T.~Graves, J.~Lowe, H.~Shah, S.~Seth, B.~Saha, C.~Curino, O.~O'Malley,
  S.~Radia, B.~Reed, and E.~Baldeschwieler.
\newblock Apache {Hadoop} {YARN}: Yet another resource negotiator.
\newblock In {\em SoCC}, 2013.

\bibitem{ernest}
S.~Venkataraman, Z.~Yang, M.~Franklin, B.~Recht, and I.~Stoica.
\newblock Ernest: efficient performance prediction for large-scale advanced
  analytics.
\newblock In {\em 13th $\{$USENIX$\}$ Symposium on Networked Systems Design and
  Implementation ($\{$NSDI$\}$ 16)}, pages 363--378, 2016.

\bibitem{borg}
A.~Verma, L.~Pedrosa, M.~Korupolu, D.~Oppenheimer, E.~Tune, and J.~Wilkes.
\newblock Large-scale cluster management at {Google} with {Borg}.
\newblock In {\em EuroSys}, 2015.

\bibitem{paris}
N.~J. Yadwadkar, B.~Hariharan, J.~E. Gonzalez, B.~Smith, and R.~H. Katz.
\newblock Selecting the best vm across multiple public clouds: a data-driven
  performance modeling approach.
\newblock In {\em Proceedings of the 2017 Symposium on Cloud Computing}, pages
  452--465. ACM, 2017.

\bibitem{delay-scheduling}
M.~Zaharia, D.~Borthakur, J.~Sen~Sarma, K.~Elmeleegy, S.~Shenker, and
  I.~Stoica.
\newblock Delay scheduling: A simple technique for achieving locality and
  fairness in cluster scheduling.
\newblock In {\em EuroSys}, 2010.

\bibitem{spark}
M.~Zaharia, M.~Chowdhury, T.~Das, A.~Dave, J.~Ma, M.~McCauley, M.~J. Franklin,
  S.~Shenker, and I.~Stoica.
\newblock {Resilient Distributed Datasets}: A fault-tolerant abstraction for
  in-memory cluster computing.
\newblock In {\em NSDI}, 2012.

\bibitem{spark-streaming}
M.~Zaharia, T.~Das, H.~Li, T.~Hunter, S.~Shenker, and I.~Stoica.
\newblock Discretized streams: Fault-tolerant stream computation at scale.
\newblock In {\em SOSP}, 2013.

\bibitem{late}
M.~Zaharia, A.~Konwinski, A.~D. Joseph, R.~Katz, and I.~Stoica.
\newblock Improving {MapReduce} performance in heterogeneous environments.
\newblock In {\em OSDI}, 2008.

\end{thebibliography}

\end{document}